\documentclass[12pt,a4paper,fontpage]{article}

\usepackage{calc,ifthen}
\usepackage{graphicx,latexsym,color}
\usepackage{amsmath,amsthm,amsfonts}
\usepackage{latexsym}
\usepackage{cite,citesort}

\usepackage{listings}

\definecolor{MyDarkBlue}{rgb}{0.1,0,0.55} 
\definecolor{MyRed}{rgb}{1,0,0} 

\usepackage{bm}
\usepackage{tikz}
\usepackage{pgfplots}
\usetikzlibrary{mindmap,trees}
\usetikzlibrary{%
  3d,
  calc,
  arrows,%
  shapes.misc,
  shapes.arrows,%
  chains,%
  matrix,%
  positioning,
  scopes,%
  decorations.pathmorphing,
  shadows%
}

\usepackage[american,siunitx]{circuitikz}
\usetikzlibrary{arrows,shapes,calc,positioning}

\renewcommand{\vec}[1]{{\boldsymbol#1}}

\newcommand{\ie}{\textit{i.e.}\/, }
\newcommand{\eg}{\textit{e.g.}\/, }
\newcommand{\cf}{\textit{cf.}\/, }

\providecommand*{\mrm}[1]{\mathrm{#1}}
\providecommand*{\unit}[1]{\ensuremath{\mrm{\,#1}}}
\providecommand*{\eu}{\ensuremath{\mrm{e}}}
\providecommand*{\iu}{\ensuremath{\mrm{i}}}

\providecommand*{\diff}{\operatorname{d}\!}
\renewcommand{\Re}{\operatorname{Re}}	
\renewcommand{\Im}{\operatorname{Im}}	

\providecommand*{\micro}{\ensuremath{\mrm{\mu}}}
\providecommand*{\degree}{\ensuremath{^\circ}}

\def\XXint#1#2#3{{\setbox0=\hbox{$#1{#2#3}{\int}$}
     \vcenter{\hbox{$#2#3$}}\kern-.5\wd0}}

\begin{document}

\title{Inductive heating of conductive nanoparticles}

\author{Sven~Nordebo\thanks{Department of Physics and Electrical Engineering, Linn\ae us University, 
351 95 V\"{a}xj\"{o}, Sweden. Phone: +46 470 70 8193. Fax: +46 470 84004. E-mail: sven.nordebo@lnu.se.},
Daniel~Sj\"{o}berg\thanks{Department of Electrical and Information Technology, Lund University, Box 118, 
221 00 Lund, Sweden. Phone: +46 46 222 7511. Fax:  +46 46 129948. E-mail:  daniel.sjoberg@eit.lth.se.},
Richard~Bayford\thanks{The Biophysics and Bioengineering Research Group, Middlesex University,
Hendon campus, The Burroughs, London, NW4 4BT, United Kingdom. E-mail: R.Bayford@mdx.ac.uk.},
}

\maketitle

\begin{abstract}
We consider the heating of biological tissue by injecting gold nanoparticles (GNPs) and subjecting the system to an electromagnetic field 
in the radio frequency spectrum. There are results that indicate that small conducting particles can substantially increase the heating locally
and thus provide a method to treat cancer.
However, recently there are also other publications that question whether metal nanoparticles can be heated in radio frequency at all.
This paper presents an analysis and some interesting observations regarding the classical electromagnetic background to this effect.
Here, it is assumed that the related dipole effects are based solely on homogeneous conducting nanospheres that are immersed in an lossy medium.
From this point of view it is concluded that the effect of using a capacitive coupling \ie a strong electric field to induce electric dipoles can be disregarded 
unless the volume fraction of the GNPs is unrealistically high, or if there are some other electric dipole mechanisms present which are not taken into account here,
such as \eg with nanospheres coated with ligands providing an electrophoretic movement and associated resonances.
On the other hand, a simplified quasi-magnetostatic analysis indicates that an inductive heating (induced eddy currents inside the metal particles) based on magnetic coupling
may have the potential to significantly increase the heating locally provided that the supplied magnetic field can be made sufficiently strong at radio frequency.  

This paper presents a near field optimization approach to study the electromagnetic heating of conductive nanoparticles.
The approach is based on a detailed analysis based on vector spherical waves for lossy materials and related energy expressions.
An optimization problem is then formulated where the power absorption inside the nanoparticles is maximized subjected to power constraints 
related to the skin effect in the surrounding medium.
The analysis shows that when the exterior medium is modelled as salty water the skin effect in the bulk material will render the simple principle of inductive heating of GNPs
practically useless at $13.56$\unit{MHz} (frequency chosen due to regulations). 
Future research will therefore be focused on an investigation of plausible dispersion mechanisms associated with coated GNPs
that can potentially generate significant absorption based on the electrophoretic resonance effects. 
\end{abstract}

\section{Introduction}

A number of publications have proposed that biological tissue can be heated quickly and selectively by the use of gold nanoparticles (GNPs)
that are subjected to a strong time-harmonic electromagnetic field, mostly at 13.56\unit{MHz} where the band is chosen due to regulations,
see \eg \cite{Collins+etal2014,Cardinal+etal2008,Gannon+etal2008,Moran+etal2009}.
The hope is to find a non-invasive method to treat cancer. 
The GNPs can also be used as a contrast agent for electrical
impedance tomography, particularly when combined with tumour targeting \cite{Callaghan+etal2010}.
The basic idea is to exploit the unique property of the cancer cells to attract the GNPs.
The rapid rate of growth of the cancer cells causes them to intake an abnormal amount of nutrients and
the GNPs can hence be coated with folic acid to target the bio-markers or antigens that are highly specific to the cancer cells, see \eg \cite{Dreaden+etal2012}.
The aim of the radio frequency (RF) treatment is then to cause local cell death only in the cancer, which in principle could be done at 
a moderate temperature of about 40 to 46 \degree C,
in addition to increasing the pore size to improve delivery of large-molecule chemotherapeutic and immunotherapeutic agents.

The physical background of the RF-heating does not seem to be fully understood and there are many phenomenological hypotheses
proposed to explain the heating \cite{Collins+etal2014,Cardinal+etal2008,Gannon+etal2008,Moran+etal2009}.
Recently, it has also been questioned whether metal nanoparticles can be heated in radio frequency at all, 
see \eg \cite{Collins+etal2014,Gupta+etal2010,Liu+etal2012,Li+etal2011,Hanson+etal2011}.

Below we comment on some key experiments and observations published 
in \cite{Collins+etal2014,Li+etal2011,Hanson+etal2011,Cardinal+etal2008,Gannon+etal2008,Moran+etal2009}.
In \cite{Moran+etal2009}, a small amount (some 10\unit{ppm} by mass) of gold is added to a 1.5\unit{mL}
container with deionized water. A strong time-varying field is applied, and the heating rate is observed to increase by factors of ten. 
It is proposed that the increased heating is due to resistive (Joule) heating and that the particle conductivity plays an important role.
A plausible explanation to this phenomenon based on classical ion transport does not seem to be provided. An argument is the nanoscale
effect associated with the electron-surface scattering where the size of the particle can be much smaller than
the mean free path of electrons in gold which is in the order of 50\unit{nm}. It can then be argued that the conductivity follows
a Drude model where the phenomenological time (or damping) constant can be seen as the typical collision time for an electron.
In small particles, it is suggested that the collisions are dominated by the boundary of the particle \cite{Link+etal1999} which will lead to collision times
in the order of femtoseconds, and which obviously can have nothing to do with properties at a few tens of\unit{MHz}.
 
In \cite{Cardinal+etal2008,Gannon+etal2008}, GNPs are injected in cancer cells and a time harmonic
electric field is applied. Increased heating and tissue destruction is observed. At the same time, there was not a significant difference in media temperatures
comparing the RF-treated cells and the control cells (no GNPs). The interpretation is that a local heat release in the microenvironment of the
cells can be sufficient to produce lethal injury to the targeted tissue.
It is difficult to read out the typical volume fraction of the nanoparticles, but it can
be expected to be very low. On the other hand, the key observation in  \cite{Gannon+etal2008} implies that maybe the temperature can be very high
locally, typically where the metal particles have clustered. This is suggested also
by some microscopic photos in \cite{Gannon+etal2008}, where the GNPs seem to have clustered
inside the cell.

It is interesting to observe that the RF-heating of gold nanoparticles is questioned and under debate.
Several authors have not been able to find theoretical nor experimental support that metal nanoparticles can be heated in radio frequency at all, 
see \eg \cite{Collins+etal2014,Gupta+etal2010,Liu+etal2012,Li+etal2011,Hanson+etal2011}.
In \cite{Hanson+etal2011} Mie scattering theory in the Rayleigh limit of very small particles is used as an argument to support
the (negative) experimental conclusions made in \cite{Li+etal2011}. 


In this paper we investigate the classical electromagnetic background to the generation of heat in a cluster of homogeneous, conductive nanoparticles that are immersed in a lossy medium.
A quasi-electrostatic homogenization approach (Hashin-Shtrikman coated spheres) is used to show that the effect of generating electric dipoles and the associated
losses dissipated in the exterior medium can be disregarded. A quasi-magnetostatic analysis is then performed
indicating that an inductive heating based on magnetic coupling may have the potential to significantly increase the heating locally provided that the 
supplied magnetic field can be made sufficiently strong at radio frequency. Finally, we present a near field optimization approach to study the electromagnetic 
heating of conductive nanoparticles  showing that when the exterior medium is modelled as salty water the skin effect in the bulk material will render the simple 
principle of inductive heating of GNPs practically useless at $13.56$\unit{MHz}. 

Future research issues are discussed in some detail, and we are particularly emphasizing the potential of exploiting the
electrophoretic resonance effects that can be associated with a suspension of glutathione coated GNPs, see \eg \cite{Nordebo+etal2016a,Sassaroli+etal2012,marquez2013hyperthermia,Callaghan+etal2010}.
Our first results in this direction are reported in \cite{Nordebo+etal2016a}.

\section{Dipole fields}

\subsection{Notation and conventions}
The following notation and conventions will be used below.
The Maxwell's equations \cite{Jackson1999} for the electric and magnetic fields $\vec{E}$ and $\vec{H}$ are considered based on SI-units 
and with time convention $\eu^{-\iu\omega t}$ for time harmonic fields.
Let $\mu_0$, $\epsilon_0$, $\eta_0$ and ${\rm c}_0$ denote the permeability, the permittivity, the wave impedance and
the speed of light in vacuum, respectively, and where $\eta_0=\sqrt{\mu_0/\epsilon_0}$ and ${\rm c}_0=1/\sqrt{\mu_0\epsilon_0}$.
The wavenumber of vacuum is given by $k_0=\omega\sqrt{\mu_0\epsilon_0}$ where $\omega=2\pi f$ is the angular frequency and $f$ the frequency.
Note that $\omega\mu_0=k_0\eta_0$ and $\omega\epsilon_0=k_0/\eta_0$. For a homogeneous and isotropic material with relative (and generally complex valued)
permeability $\mu$ and permittivity $\epsilon$ expressed in the frequency domain, the corresponding wavenumber
and wave impedance are given by $k=k_0\sqrt{\mu\epsilon}$ and $\eta_0\eta$, respectively, and where the relative wave impedance is $\eta=\sqrt{\mu/\epsilon}$.
Note that $\omega\mu_0\mu=k\eta_0\eta$ and $\omega\epsilon_0\epsilon=k/\eta_0\eta$. 
For a conductive material the conductivity is denoted $\sigma$ and the corresponding complex valued relative permittivity is given by
$\epsilon=\epsilon_{\rm r}+\iu \sigma/\omega\epsilon_0$ where $\epsilon_{\rm r}$ is the real valued relative permittivity.
The spherical coordinates are denoted by $(r,\theta,\phi)$, the corresponding unit vectors $(\hat{\vec{r}},\hat{\vec{\theta}},\hat{\vec{\phi}})$,
and the radius vector $\vec{r}=r\hat{\vec{r}}$.  The cartesian unit vectors are denoted $(\hat{\vec{x}},\hat{\vec{y}},\hat{\vec{z}})$.

\subsection{Electric and magnetic dipole fields}

Following the definition of the vector spherical waves given in Appendix \ref{sect:spherical} where the general multipole indices are $(\tau,m,l)$,
the electromagnetic fields of a transverse electric (TE) magnetic dipole ($\tau=1$) and a transverse magnetic (TM) electric dipole ($\tau=2$)
with dipole moments in the $\hat{\vec{z}}$-direction, \ie $(m,l)=(0,1)$, can be expressed as follows, \cf also \cite{Bostrom+Kristensson+Strom1991,Arfken+Weber2001,Jackson1999,Newton2002}.
The regular dipole waves are
\begin{equation}\label{eq:regdipoles}
\left\{\begin{array}{l}
\displaystyle \vec{E}(\vec{r})=a_{\rm M}{{\rm j}_1(kr)}\hat{\vec{\phi}}\sin\theta+a_{\rm E}\left(-\left(\frac{{\rm j}_1(kr)}{kr}+{\rm j}_1^{\prime}(kr) \right)\hat{\vec{\theta}}\sin\theta
+\frac{{\rm j}_1(kr)}{kr}2\hat{\vec{r}}\cos\theta\right), \vspace{0.2cm} \\
\displaystyle \iu\eta_0\eta\vec{H}(\vec{r})=a_{\rm M}\left(-\left(\frac{{\rm j}_1(kr)}{kr}+{\rm j}_1^{\prime}(kr) \right)\hat{\vec{\theta}}\sin\theta
+\frac{{\rm j}_1(kr)}{kr}2\hat{\vec{r}}\cos\theta\right)+a_{\rm E}{{\rm j}_1(kr)}\hat{\vec{\phi}}\sin\theta,
\end{array}\right.
\end{equation}
and the outgoing dipole waves
\begin{equation}\label{eq:singdipoles}
\left\{\begin{array}{l}
\displaystyle \vec{E}(\vec{r})=b_{\rm M}{{\rm h}_1^{(1)}(kr)}\hat{\vec{\phi}}\sin\theta+b_{\rm E}\left(-\left(\frac{{\rm h}_1^{(1)}(kr)}{kr}+{{\rm h}_1^{(1)}}^{\prime}(kr) \right)\hat{\vec{\theta}}\sin\theta
+\frac{{\rm h}_1^{(1)}(kr)}{kr}2\hat{\vec{r}}\cos\theta\right), \vspace{0.2cm} \\
\displaystyle \iu\eta_0\eta\vec{H}(\vec{r})=b_{\rm M}\left(-\left(\frac{{\rm h}_1^{(1)}(kr)}{kr}+{{\rm h}_1^{(1)}}^{\prime}(kr) \right)\hat{\vec{\theta}}\sin\theta
+\frac{{\rm h}_1^{(1)}(kr)}{kr}2\hat{\vec{r}}\cos\theta\right)+b_{\rm E}{{\rm h}_1^{(1)}(kr)}\hat{\vec{\phi}}\sin\theta,
\end{array}\right.
\end{equation}
and where the dipole coefficients are $a_{\rm M}=a_{101}\sqrt{3/8\pi}$, $a_{\rm E}=a_{201}\sqrt{3/8\pi}$, $b_{\rm M}=b_{101}\sqrt{3/8\pi}$ and $b_{\rm E}=b_{201}\sqrt{3/8\pi}$
as defined in Appendix \ref{sect:spherical}. Here, ${\rm j}_1(x)$ and ${\rm h}_1^{(1)}(x)$ are the spherical Bessel function and the spherical Hankel function of the first kind, respectively,
both of order $l=1$.

Consider now the small argument asymptotics of ${\rm j}_1(x)$ and ${\rm h}_1^{(1)}(x)$
\begin{equation}
\left\{\begin{array}{l}
{\rm j}_1(x)=\displaystyle \frac{\sin x}{x^2}-\frac{\cos x}{x}\sim\frac{1}{3}x, \vspace{0.2cm} \\
{\rm h}_1^{(1)}(x)=-\displaystyle\frac{\eu^{\iu x}}{x}\left( 1+\frac{\iu}{x}\right)\sim -\iu\frac{1}{x^2},
\end{array}\right.
\end{equation}
which are valid as $|x|\rightarrow 0$. The corresponding quasi-static regular dipole fields are
\begin{equation}\label{eq:regdipolesassymptotic}
\left\{\begin{array}{l}
\displaystyle \vec{E}(\vec{r})=a_{\rm M}kr\hat{\vec{\phi}}\sin\theta+a_{\rm E}2\left(\hat{\vec{r}}\cos\theta-\hat{\vec{\theta}}\sin\theta \right), \vspace{0.2cm} \\
\displaystyle \iu\eta_0\eta\vec{H}(\vec{r})=a_{\rm M}2\left(\hat{\vec{r}}\cos\theta-\hat{\vec{\theta}}\sin\theta \right)+a_{\rm E}kr\hat{\vec{\phi}}\sin\theta,
\end{array}\right.
\end{equation}
where the factor $1/3$ has been absorbed in the coefficients $a_{\rm M}$ and $a_{\rm E}$.
Similarly, the corresponding quasi-static singular dipole fields  are
\begin{equation}\label{eq:singdipolesassymptotic}
\left\{\begin{array}{l}
\displaystyle \vec{E}(\vec{r})=b_{\rm M}\frac{1}{(kr)^2}\hat{\vec{\phi}}\sin\theta+b_{\rm E}\frac{1}{(kr)^3}\left(2\hat{\vec{r}}\cos\theta+\hat{\vec{\theta}}\sin\theta \right), \vspace{0.2cm} \\
\displaystyle \iu\eta_0\eta\vec{H}(\vec{r})=b_{\rm M}\frac{1}{(kr)^3}\left(2\hat{\vec{r}}\cos\theta+\hat{\vec{\theta}}\sin\theta \right)+b_{\rm E}\frac{1}{(kr)^2}\hat{\vec{\phi}}\sin\theta,
\end{array}\right.
\end{equation}
where the factor $-\iu$ has been absorbed in the coefficients $b_{\rm M}$ and $b_{\rm E}$.
In \eqref{eq:regdipolesassymptotic}, it is noted that $\hat{\vec{r}}\cos\theta-\hat{\vec{\theta}}\sin\theta=\hat{\vec{z}}$ which represents the direction
of the dipole moment. The term ``quasi-static'' is employed above due to the fact that the representations \eqref{eq:regdipolesassymptotic} and \eqref{eq:singdipolesassymptotic} are
valid as $|kr|\rightarrow 0$, \ie in the low-frequency (or long wavelength) limit, as well as when the physical dimensions of the particles (as with nanoparticles) become very
small for any fixed frequency.

\section{Quasi-electrostatic heating}\label{sect:quasielectrostatic}

We investigate first the potential of electromagnetic heating based on the quasi-electrostatic phenomenon, \ie a polarization of the nanostructure based on the generation of electric dipoles.
In this case, there will be very small electric fields inside the highly conducting spheres and the heating will take place due to a strong electric field in the surrounding medium.

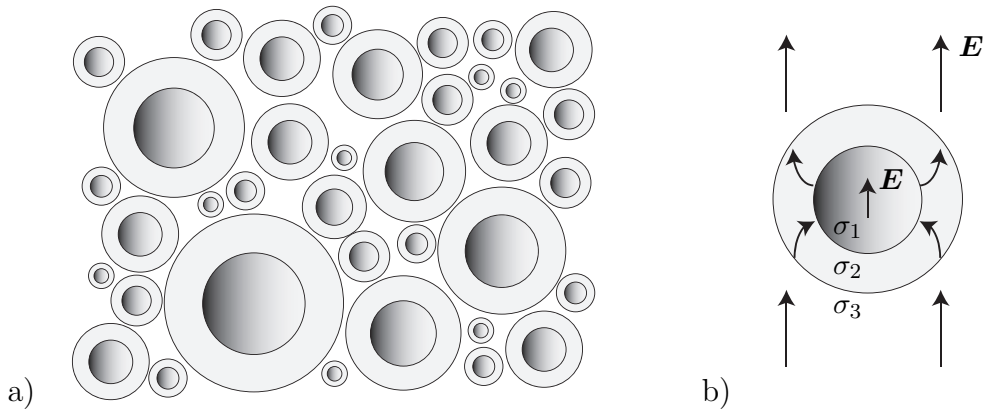
\begin{figure}[htb]
\begin{picture}(50,150)
\put(80,0){\makebox(150,150){\includegraphics[width=7cm]{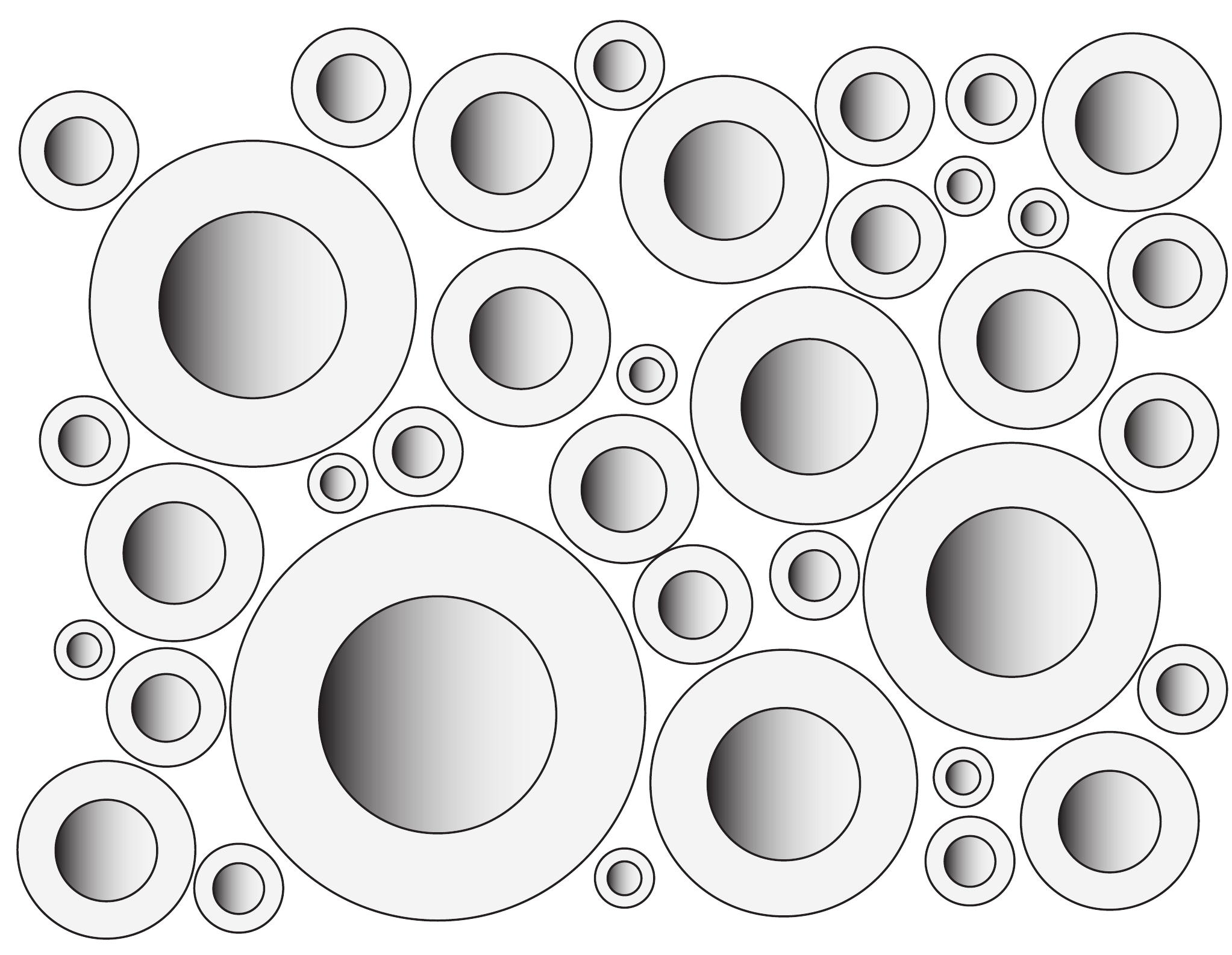}}} 
\put(280,0){\makebox(150,150){\includegraphics[width=3cm]{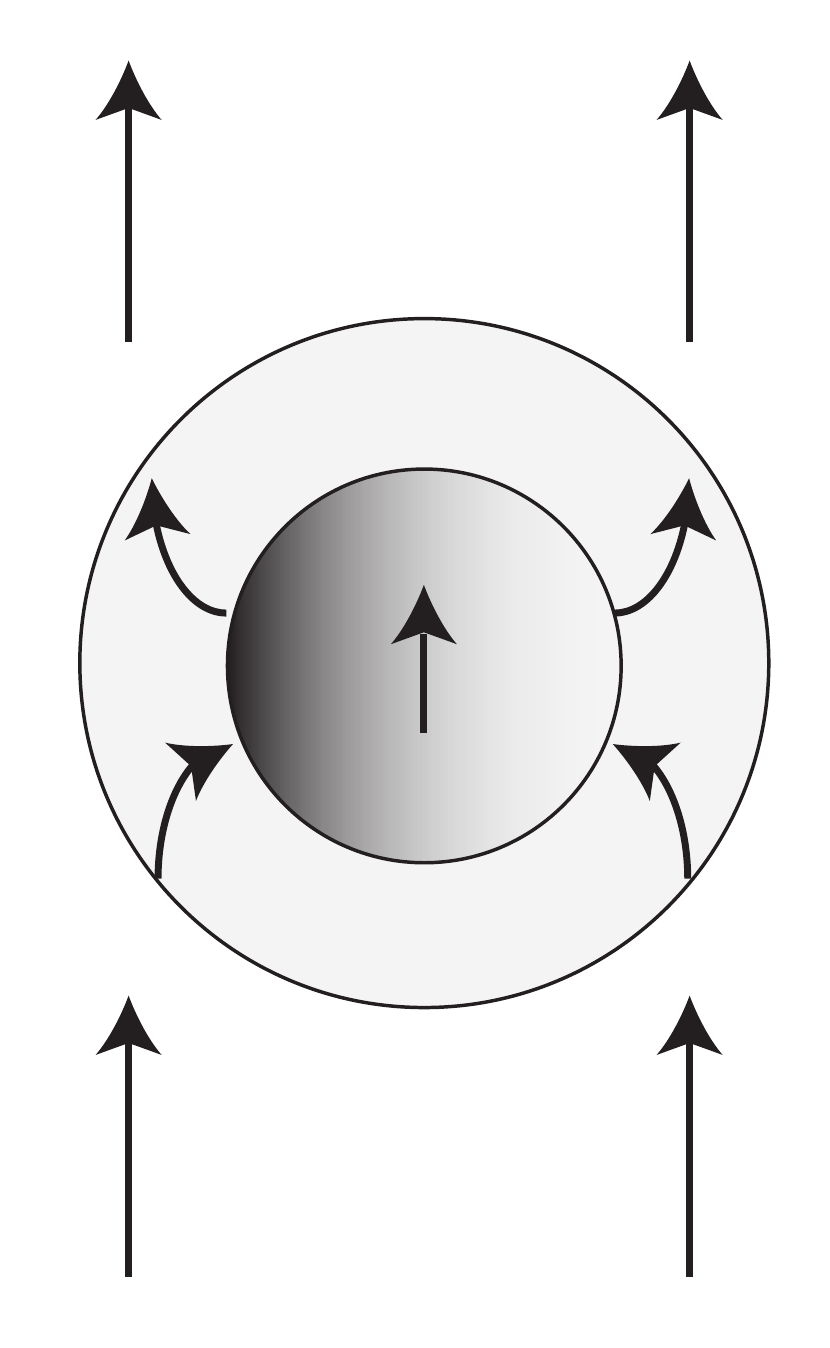}}} 
\put(30,0){ a)}
\put(290,0){ b)}
\put(360,80){\small $\vec{E}$}
\put(390,130){\small $\vec{E}$}
\put(343,63){\small $\sigma_1$}
\put(343,48){\small $\sigma_2$}
\put(343,33){\small $\sigma_3$}
\end{picture}
\caption{a) Geometry of the Hashin-Shtrikman coated spheres.
b) Unit electric dipole for the quasi-electrostatic approximation.}
\label{fig:HSspheres}
\end{figure}

To analyze this situation we consider the classical homogenization technique based on the Hashin-Shtrikman coated spheres assemblage \cite{Milton2002} as depicted in Fig.~\ref{fig:HSspheres}.
The spheres are of different size with a fixed volume fraction $f_1=r_1^3/r_2^3$ where $r_1$ is the radius of the inner sphere having conductivity $\sigma_1$
and $r_2$ the outer radius of the coating having conductivity $\sigma_2$ and which also represents the background medium, see Figure~\ref{fig:HSspheres}b. 
The fictitious exterior region outside the coated spheres has a conductivity $\sigma_3$ which is chosen so that the spheres are ``cloaked'' for a homogeneous
externally applied electric field $\vec{E}$ as depicted in Figure~\ref{fig:HSspheres}b. This means that the coated spheres do not have a dipole moment when placed in the 
fictitious exterior medium. The fictitious exterior medium can therefore be replaced by coated spheres of different size filling the whole space 
without affecting the externally applied field and in this way the procedure provides a solution to the corresponding homogenization problem.

Based on \eqref{eq:regdipolesassymptotic} and \eqref{eq:singdipolesassymptotic}, the quasi-static electric dipole field can be written
\begin{equation}\label{eq:Efield}
\vec{E}(\vec{r})=\left\{\begin{array}{ll}
a_1E_0\left(\hat{\vec{r}}\cos\theta-\hat{\vec{\theta}}\sin\theta \right) & r<r_1, \vspace{0.2cm} \\ 
\displaystyle a_2E_0\left(\hat{\vec{r}}\cos\theta-\hat{\vec{\theta}}\sin\theta \right) 
+b_2E_0\frac{1}{r^3}\left(2\hat{\vec{r}}\cos\theta+\hat{\vec{\theta}}\sin\theta \right)
& r_1<r<r_2, \vspace{0.2cm} \\
E_0\left(\hat{\vec{r}}\cos\theta-\hat{\vec{\theta}}\sin\theta \right) & r_2<r,
\end{array}\right.
\end{equation}
where the externally applied electric field is $E_0\hat{\vec{z}}$. 
The appropriate boundary conditions are given by the continuity of the tangential electric field $\vec{E}$ and of the normal component of the
current density $\vec{J}=\sigma\vec{E}$ (divergence free currents) at $r_1$ and $r_2$, respectively. This yields
\begin{equation}
\left\{\begin{array}{l}
\displaystyle a_1E_0=a_2E_0-b_2E_0\frac{1}{r_1^3},  \vspace{0.2cm} \\ 
\displaystyle a_2E_0-b_2E_0\frac{1}{r_2^3}=E_0,  \vspace{0.2cm} \\ 
\displaystyle \sigma_1a_1E_0=\sigma_2\left(a_2E_0+2b_2E_0\frac{1}{r_1^3} \right),\vspace{0.2cm} \\
\displaystyle \sigma_2\left(a_2E_0+2b_2E_0\frac{1}{r_2^3} \right)=\sigma_3E_0,
\end{array}\right.
\end{equation}
with the unique solution
\begin{equation}\label{eq:solution}
\left\{\begin{array}{l}
\displaystyle a_1=\frac{3}{3+f_2\left(\sigma_1/\sigma_2-1 \right)}, \vspace{0.2cm} \\ 
\displaystyle a_2=1+\frac{f_1\left(\sigma_1/\sigma_2-1 \right)}{3+f_2\left(\sigma_1/\sigma_2-1 \right)}, \vspace{0.2cm} \\ 
\displaystyle b_2=\frac{r_1^3\left(\sigma_1/\sigma_2-1 \right)}{3+f_2\left(\sigma_1/\sigma_2-1 \right)}, \vspace{0.2cm} \\ 
\displaystyle \sigma_3=\sigma_2\left[1+ \frac{3f_1\left(\sigma_1/\sigma_2-1 \right)}{3+f_2\left(\sigma_1/\sigma_2-1 \right)} \right],
\end{array}\right.
\end{equation}
and where $f_1=r_1^3/r_2^3$ is the volume fraction of conductive nanoparticles and $f_2=1-f_1$. 
The solution for $\sigma_3$ is the well known Hashin-Shtrikman/Maxwell-Garnet homogenization formula\cite{Milton2002}.

The power in the inner sphere is given by
\begin{equation}
S_1^{\rm e}=\frac{1}{2}\int_{r=0}^{r_1}\sigma_1|E_0|^2|a_1|^2\diff v=\sigma_2\frac{|E_0|^2}{2}\frac{4\pi r_1^3}{3}\frac{9\sigma_1/\sigma_2}{\left| 3+f_2(\sigma_1/\sigma_2-1) \right|^2},
\end{equation}
and which will tend to zero as $\sigma_1/\sigma_2\rightarrow \infty$.

It can be shown that the power in the coating is given by
\begin{multline}
S_2^{\rm e}=\frac{1}{2}\int_{r=r_1}^{r_2}\sigma_2|E_0|^2
\left|a_2\left(\hat{\vec{r}}\cos\theta-\hat{\vec{\theta}}\sin\theta \right) 
+b_2\frac{1}{r^3}\left(2\hat{\vec{r}}\cos\theta+\hat{\vec{\theta}}\sin\theta \right)\right|^2\diff v \\
=\frac{1}{2}\sigma_2|E_0|^2\frac{4\pi}{3}\left[|a_2|^2(r_2^3-r_1^3)+2|b_2|^2\left( \frac{1}{r_1^3}-\frac{1}{r_2^3}\right) \right],
\end{multline}
where we have employed $\diff v=r^2\sin\theta\diff r\diff\theta\diff\phi$, $\int_{0}^{\pi}\cos^2\theta\sin\theta\diff\theta=\frac{2}{3}$ and
$\int_{0}^{\pi}\sin^3\theta\diff\theta=\frac{4}{3}$.

By using the asymptotics $a_2\sim 1+\frac{f_1}{f_2}$ and $b_2\sim r_1^3\frac{1}{f_2}$ which are valid as $\sigma_1/\sigma_2\rightarrow \infty$, 
it is found that
\begin{equation}
S_2^{\rm e}\sim\frac{1}{2}\sigma_2|E_0|^2\frac{4\pi}{3}r_2^{3}\left(1+\frac{3f_1}{f_2} \right),
\end{equation}
as $\sigma_1/\sigma_2\rightarrow \infty$.
Using as a reference the background heating
\begin{equation}
S_0^{\rm e}=\frac{1}{2}\sigma_2|E_0|^2\frac{4\pi}{3}r_2^{3},
\end{equation}
the relative heating coefficient is defined as
\begin{equation}\label{eq:Fedef}
F^{\rm e}=\lim_{\sigma_1/\sigma_2\rightarrow \infty}\frac{S_1^{\rm e}+S_2^{\rm e}-S_0^{\rm e}}{S_0^{\rm e}}=\frac{3f_1}{1-f_1},
\end{equation}
where $f_1=r_1^3/r_2^3$ is the volume fraction of conductive nanospheres.

\section{Quasi-magnetostatic heating}\label{sect:quasimagnetostatic}

Next we investigate the potential of electromagnetic heating based on the quasi-magnetostatic phenomenon, \ie a polarization of the nanostructure based on the generation of magnetic dipoles.
In this case, there will be a very strong magnetic field penetrating the small spheres and the heating will take place due to the induced eddy currents inside the 
highly conducting spheres. To motivate this assumption, it is noted that the skin-depth in gold is 75\unit{\micro m} at 1\unit{MHz} and the size of a gold nanoparticle is typically 
in the order of 5\unit{nm} \cite{Gannon+etal2008}. 
Hence, the simplified analysis below is based on the assumption that the self-inductance as well as the mutual inductances of the spheres can be neglected 
and that the magnetic field generated by the induced currents will have a very little effect on the externally applied magnetic field. This is a common
assumption that can be applied in magneto-statics when the skin-depth of the material is much larger than the physical dimensions of the structure, 
see \eg \cite{Jackson1999,Nordebo+Gustafsson2014a}.

The analysis technique based on the Hashin-Shtrikman coated spheres that was used above can not be
readily applied to analyze the case with inductive heating. As will be seen below, the reason for this is simply the fact that the 
actual power loss in each coated sphere depends on its radius and it is hence difficult to assess the correct background heating
when there are no spheres present. Note also that we have made no particular assumptions about the externally applied electromagnetic 
field (plane, cylindrical or spherical waves, etc) other than that the magnetic field behaves locally as a quasi-static homogeneous field. 

\begin{figure}[htb]
\begin{picture}(50,150)
\put(80,0){\makebox(150,150){\includegraphics[width=8.5cm]{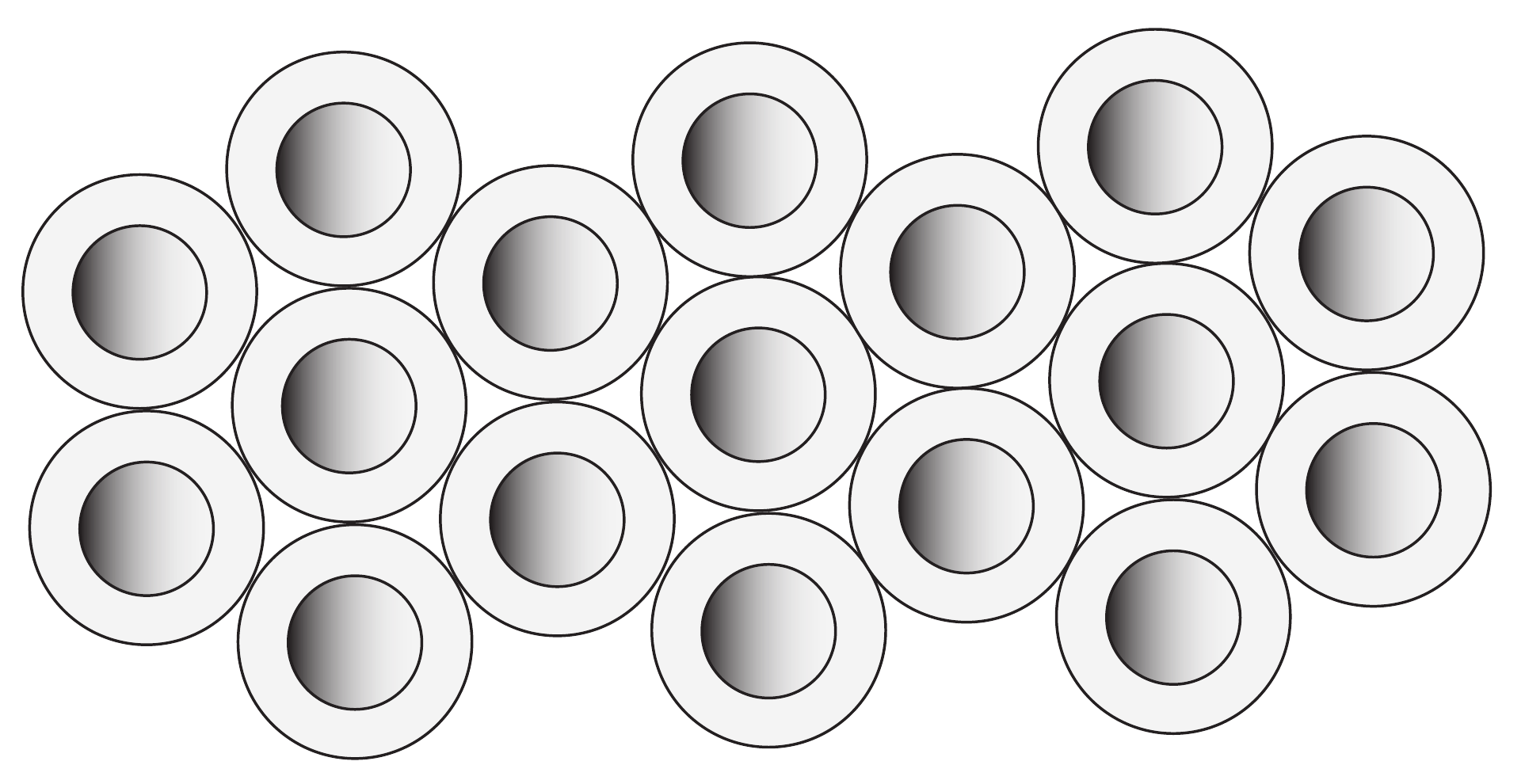}}} 
\put(280,0){\makebox(150,150){\includegraphics[width=3cm]{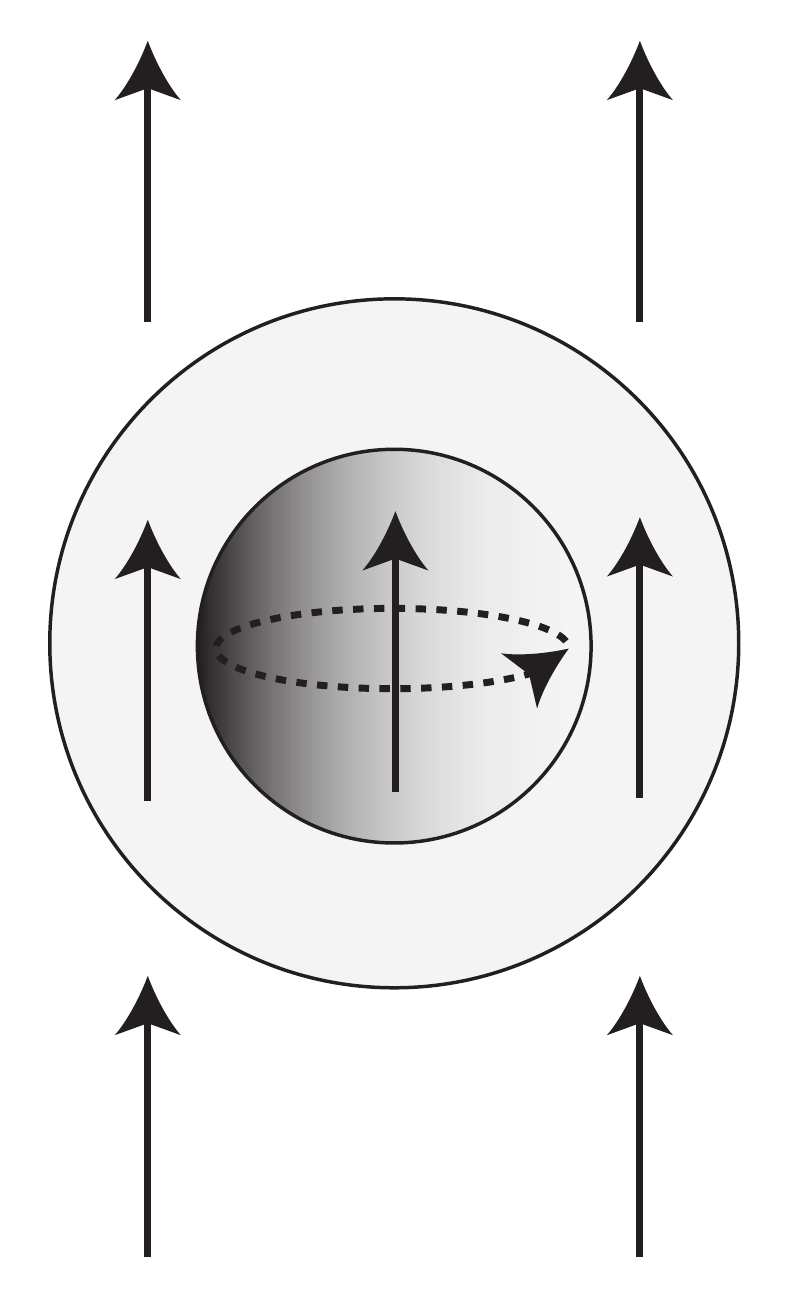}}} 
\put(30,0){ a)}
\put(290,0){ b)}
\put(358,82){\small $\vec{H}$}
\put(388,132){\small $\vec{H}$}
\put(360,61){\small $\vec{E}$}
\put(343,60){\small $\sigma_1$}
\put(343,45){\small $\sigma_2$}
\end{picture}
\caption{a) Hexagonal closed packed nanospheres.
b) Unit magnetic dipole for the quasi-magnetostatic approximation.}
\label{fig:HCPspheres}
\end{figure}
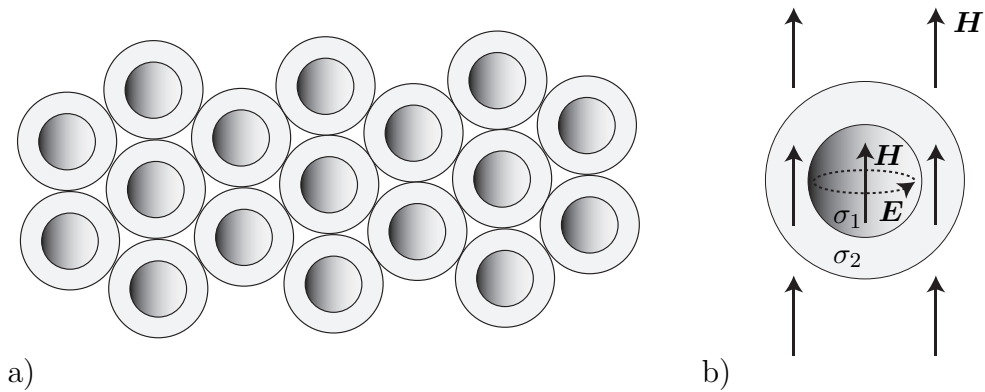

To analyze the situation with inductive heating we consider therefore a periodic structure of equal spheres as indicated in Figure \ref{fig:HCPspheres}.
This is a realistic assumption since the nanoparticles can usually be processed to have almost equal size.
We emphasize however that this analysis is simplified in the sense that we are assuming that the actual nanostructure is large enough to be considered periodic
at the same time as it is small enough to justify that the skin effect can be neglected.
It is also noted that the highest fraction of space occupied by a lattice of equal spheres is about 74\% and is achieved \eg with a hexagonal closed packed (hcp) or
a face centered cubic (fcc) structure \cite{Kittel1986}.
In the simplified analysis that is executed below we simply disregard the remaining 26\% which may also contribute to the heating.
Each cell in the lattice is therefore described as a coated sphere where $r_1$ is the radius of the inner sphere having conductivity $\sigma_1$
and $r_2$ the outer radius of the coating having conductivity $\sigma_2$ and which also represents the background medium, see Figure~\ref{fig:HCPspheres}b. 

Based on \eqref{eq:regdipolesassymptotic}, the quasistatic magnetic dipole field inside the coated sphere can be written
\begin{equation}\label{eq:Hfield}
\vec{H}(\vec{r})=H_0\left(\hat{\vec{r}}\cos\theta-\hat{\vec{\theta}}\sin\theta \right) \quad r<r_2,
\end{equation}
and the induced electric field can be expressed as
\begin{equation}\label{eq:indEfield}
\vec{E}(\vec{r})=\iu\omega\mu_0 H_0\hat{\vec{\phi}}\frac{1}{2}r\sin\theta \quad r<r_2,
\end{equation}
in accordance to the Faraday's law of induction $\nabla\times\vec{E}(\vec{r})=\iu\omega\mu_0\vec{H}(\vec{r})$.


The power generated inside the sphere of radius $r_1$ is given by
\begin{multline}\label{eq:S1mdef}
S_1^{\rm m}=\frac{1}{2}\int_{r=0}^{r_1}\sigma_1|\vec{E}(\vec{r})|^2\diff v 
=\frac{1}{2}\sigma_1\left|\omega\mu_0H_0\frac{1}{2}\right|^2\int_0^{r_1}r^4\diff r\int_{0}^{\pi}\sin^3\theta\diff\theta\int_0^{2\pi}\diff\phi \\
=\pi\sigma_1|\omega\mu_0H_0|^2\frac{1}{15}r_1^5,
\end{multline}
where we have used $\diff v=r^2\sin\theta\diff r\diff\theta\diff\phi$ and $\int_0^\pi\sin^3\theta\diff\theta=4/3$. 
Similarly, the reference case with a homogeneous background without the nanoparticles is given by
\begin{equation}\label{eq:S0mdef}
S_0^{\rm m}=\frac{1}{2}\int_{r=0}^{r_2}\sigma_2|\vec{E}(\vec{r})|^2\diff v 
=\pi\sigma_2|\omega\mu_0H_0|^2\frac{1}{15}r_2^5.
\end{equation}
The power generated in the spherical region between radius $r_1$ and $r_2$ is given by
\begin{multline}\label{eq:S2mdef}
S_2^{\rm m}=\frac{1}{2}\int_{r=r_1}^{r_2}\sigma_2|\vec{E}(\vec{r})|^2\diff v 
=\frac{1}{2}\sigma_2\left|\omega\mu_0H_0\frac{1}{2}\right|^2\int_{r_1}^{r_2}r^4\diff r\int_{0}^{\pi}\sin^3\theta\diff\theta\int_0^{2\pi}\diff\phi \\
=\pi\sigma_2|\omega\mu_0H_0|^2\frac{1}{15}\left(r_2^5-r_1^5\right).
\end{multline}

The increase in power loss relative to the background power loss without the nanoparticles is given by the relative heating coefficient
\begin{equation}\label{eq:Fmdef}
F^{\rm m}=\frac{S_1^{\rm m}+S_2^{\rm m}-S_0^{\rm m}}{S_0^{\rm m}}=\frac{\sigma_1-\sigma_2}{\sigma_2}f_1^{5/3},
\end{equation}
where $f_1=r_1^3/r_2^3$ is the volume fraction of conductive nanoparticles.
The power per unit volume (in\unit{W/m^3}) for the reference case is
\begin{equation}\label{eq:P0def}
P_0=\frac{S_0^{\rm m}}{4\pi r_2^3/3}=\sigma_2\omega^2B_0^2\frac{r_2^2}{20}=\sigma_2\omega^2B_0^2\frac{r_1^2}{20}\left(\frac{1}{f_1} \right)^{2/3},
\end{equation}
where $B_0=\mu_0H_0$. Finally, the local inductive heating (in\unit{W/m^3}) is given by
\begin{equation}\label{eq:Ploc}
P_{\rm loc}=F^{\rm m}P_0=(\sigma_1-\sigma_2)\omega^2B_0^2\frac{r_1^2}{20}f_1.
\end{equation}

\section{Mie theory for small conducting spheres}\label{sect:Mie}
Mie theory can be used to investigate the extinction cross section of a small conducting sphere in a lossless host medium, see \eg \cite{Shore2015,Hanson+etal2011}.
Let the nanospheres and the exterior region have normalized dielectric constants $\epsilon_1$ and $\epsilon_2$, respectively, and where
$\epsilon_2$ is assumed to be real valued for a lossless exterior region. 
The scattering and the extinction cross sections $\sigma_{\rm s}$ and $\sigma_{\rm ext}$ are given by
\begin{equation}\label{eq:signmas}
\sigma_{\rm s}=\frac{2\pi}{k^2}\sum_{l=1}^\infty\left( 2l+1\right)\left\{ |t^{\rm E}_l|^2+|t^{\rm M}_l|^2 \right\},
\end{equation}
and
\begin{equation}\label{eq:signmaext}
\sigma_{\rm ext}=-\frac{2\pi}{k^2}\sum_{l=1}^\infty\left( 2l+1\right)\Re\left\{ t^{\rm E}_l+t^{\rm M}_l\right\},
\end{equation}
where $k=k_0\sqrt{\epsilon_2}$ is the wave number of the exterior region and where
$t^{\rm E}_l$ and $t^{\rm M}_l$ are the electric and the magnetic Mie coefficients which are based on the scattering of vector spherical waves
due to an impinging plane wave, see \eg \cite{Shore2015}.

To the lowest order in $kr_1$, the dominating Mie coefficients for the nonmagnetic conducting sphere are given by
\begin{equation}
\begin{array}{l}
\displaystyle t^{\rm E}_1=\iu2\frac{(kr_1)^3}{3}\frac{\epsilon_1-\epsilon_2}{\epsilon_1+2\epsilon_2}, \vspace{0.2cm} \\
\displaystyle t^{\rm M}_1=\iu\frac{(kr_1)^5}{45}\frac{\epsilon_1-\epsilon_2}{\epsilon_2}.
\end{array}
\end{equation}
The Rayleigh approximation for small particles can hence be written
\begin{equation}
\sigma_{\rm ext}=k\Im\left\{ \alpha^{\rm E}+\alpha^{\rm M} \right\},
\end{equation}
where the electric and the magnetic polarizability are, respectively
\begin{equation}
\begin{array}{l}
\displaystyle \alpha^{\rm E}=4\pi r_1^3\frac{\epsilon_1-\epsilon_2}{\epsilon_1+2\epsilon_2}, \vspace{0.2cm} \\
\displaystyle \alpha^{\rm M}=4\pi r_1^3\frac{(kr_1)^2}{30}\frac{\epsilon_1-\epsilon_2}{\epsilon_2}.
\end{array}
\end{equation}

The nanospheres are assumed to have conductivity $\sigma_1$ and hence $\epsilon_1=1+\iu \sigma_1/\omega\epsilon_0$
where the real part of the permittivity can be assumed to be unity for most metals \cite{Cheng1989}.
Further, let $\epsilon_2=1$ corresponding to a host medium consisting of vacuum (or air).
Then
\begin{equation}
\begin{array}{l}
\displaystyle \alpha^{\rm E}=4\pi r_1^3\frac{\iu\sigma_1}{\iu\sigma_1+3\omega\epsilon_0}, \vspace{0.2cm} \\
\displaystyle \alpha^{\rm M}=4\pi r_1^3\frac{(kr_1)^2}{30}\frac{\iu\sigma_1}{\omega\epsilon_0},
\end{array}
\end{equation}
and where  $\alpha^{\rm E}\sim 4\pi r_1^3$ in the high-conductivity limit where $\sigma_1\rightarrow\infty$.
Neglecting the contribution from the electric dipole the extinction cross-section of the conducting sphere becomes
\begin{equation}\label{eq:sigmaextM}
\sigma_{\rm ext}^{\rm M}=4\pi r_1^3\frac{(kr_1)^2}{30}\sigma_1\eta_0,
\end{equation}
which is due to the induced magnetic dipole moment (induced eddy currents).

Neglecting the multiple scattering effect that is associated with the cluster of nanoparticles,
the local inductive heating (in\unit{W/m^3}) is given by
\begin{equation}
P_{\rm loc}=n\sigma_{\rm ext}^{\rm M}\frac{\left| E_0\right|^2}{2\eta_0},
\end{equation}
where $E_0$ is the amplitude of the incident electric field and $n$ the number of particles per unit volume given by
\begin{equation}
n=\frac{f_1}{4\pi r_1^3/3},
\end{equation}
and where $f_1$ is the volume fraction of nanoparticles.
Finally, the  local inductive heating becomes
\begin{equation}
P_{\rm loc}=\left| E_0 \right|^2f_1\frac{(kr_1)^2}{20}\sigma_1,
\end{equation}
or 
\begin{equation}\label{eq:Ploc2}
P_{\rm loc}=\sigma_1\omega^2B_0^2\frac{r_1^2}{20}f_1,
\end{equation}
where $E_0=\eta_0B_0/\mu_0$ and $B_0$ is the amplitude of the magnetic flux density.

 It should be noted that the Mie theory above is based on the assumption that the incident field is a plane wave in a lossless medium.
 It is also noted that the result \eqref{eq:Ploc2} above when interpreted in terms of the magnetic flux density $B_0$
 agrees very well with the result \eqref{eq:Ploc} which is based on the quasi-magnetostatic approximation.
 

\section{Optimal near field and skin effect}

%
%
%

Consider the scattering of electromagnetic waves from a single conducting nanosphere with radius $r_1$ (region 1) and complex valued relative permittivity 
\begin{equation}\label{eq:epsilon1def}
\epsilon_1(\omega)=1+\iu\frac{\sigma_1}{\omega\epsilon_0},
\end{equation} 
where $\sigma_1$ is the conductivity of the nanosphere. The wavenumber of the material inside the spherical region 1 is given by $k_1=k_0\sqrt{\epsilon_1}$.

The small conducting sphere is centered inside a larger sphere of radius $r_2$ (region 2) where the
exterior material is considered to be homogeneous. 
The exterior material is modelled  by using a combined Debye and a conductivity model (as with salty water)
with complex valued relative permittivity
\begin{equation}\label{eq:epsilondef}
\epsilon(\omega)=\epsilon_{\infty}+\frac{\epsilon_{\rm s}-\epsilon_{\infty}}{1-\iu\omega\tau}+\iu\frac{\sigma}{\omega\epsilon_0},
\end{equation}
where $\epsilon_{\infty}$, $\epsilon_{\rm s}$ and $\tau$ are the high frequency permittivity,  the static permittivity and the relaxation time of the Debye model, respectively,
and $\sigma$ is the conductivity. The wavenumber of the exterior material is given by $k=k_0\sqrt{\epsilon}$.

Consider a description of the electromagnetic field based on an expansion in vector spherical waves as defined in \eqref{eq:Esphdef}.
The applied external field is generated by arbitrary sources outside the sphere of radius $r_2$ and
expanded in the region $r_1<r<r_2$ by using regular spherical waves with multipole coefficients $a_{\tau ml}^{(2)}$.
The scattered field for $r>r_1$ is given by the multipole coefficients 
\begin{equation}
b_{\tau ml}^{(2)}=t_{\tau l}a_{\tau ml}^{(2)},
\end{equation}
where $t_{\tau l}$ is the transition matrix generating the exterior fields.
The field inside the nanosphere for $r<r_1$ is given by the multipole coefficients 
\begin{equation}
a_{\tau ml}^{(1)}=r_{\tau l}a_{\tau ml}^{(2)},
\end{equation}
where $r_{\tau l}$ is the transition matrix generating the interior fields.
The transition matrices\footnote{In section \ref{sect:Mie}, $t_{1 l}$ and $t_{2 l}$ are denoted $t^{\rm M}_l$ and $t^{\rm E}_l$, respectively.}
$t_{\tau l}$ and $r_{\tau l}$ are readily obtained by applying the appropriate boundary conditions
based on the tangential electric and magnetic fields. It can be readily shown that
\begin{equation} 
\left\{\begin{array}{l}
t_{1l}=   \displaystyle \frac{\textrm{j}_l(kr_1) (k_1r_1\textrm{j}_l(k_1r_1))^\prime \mu -\textrm{j}_l(k_1r_1) (kr_1\textrm{j}_l(kr_1))^\prime \mu_1}
{\textrm{j}_l(k_1r_1) (kr_1\textrm{h}_l^{(1)}(kr_1))^\prime \mu_1-\textrm{h}_l^{(1)}(kr_1) (k_1r_1\textrm{j}_l(k_1r_1))^\prime \mu },
\vspace{0.2cm} \\
t_{2l} =   \displaystyle  \frac{ \textrm{j}_l(k_1r_1) (kr_1\textrm{j}_l(kr_1))^\prime \epsilon_1 - \textrm{j}_l(kr_1) (k_1r_1\textrm{j}_l(k_1r_1))^\prime \epsilon}
{ \textrm{h}_l^{(1)}(kr_1) (k_1r_1\textrm{j}_l(k_1r_1))^\prime\epsilon- \textrm{j}_l(k_1r_1) (kr_1\textrm{h}_l^{(1)}(kr_1))^\prime\epsilon_1},
\vspace{0.2cm} \\
r_{1l} = \displaystyle  \frac{\textrm{j}_l(kr_1) (kr_1\textrm{h}_l^{(1)}(kr_1))^\prime \mu_1-\textrm{h}_l^{(1)}(kr_1) (kr_1\textrm{j}_l(kr_1))^\prime \mu_1}
{\textrm{j}_l(k_1r_1) (kr_1\textrm{h}_l^{(1)}(kr_1))^\prime \mu_1-\textrm{h}_l^{(1)}(kr_1) (k_1r_1\textrm{j}_l(k_1r_1))^\prime \mu },
\vspace{0.2cm} \\
r_{2l} =  \displaystyle  -\frac{ (\textrm{j}_l(kr_1) (kr_1\textrm{h}_l^{(1)}(kr_1))^\prime-\textrm{h}_l^{(1)}(kr_1) (kr_1\textrm{j}_l(kr_1))^\prime)\sqrt{\epsilon} \sqrt{\epsilon_1} \sqrt{\mu_1}}
   { ( \textrm{h}_l^{(1)}(kr_1) (k_1r_1\textrm{j}_l(k_1r_1))^\prime\epsilon- \textrm{j}_l(k_1r_1) (kr_1\textrm{h}_l^{(1)}(kr_1))^\prime\epsilon_1)\sqrt{\mu }},
   \end{array}\right.
\end{equation}
see Appendix \ref{sect:spherical} for the definitions of the spherical Bessel and Hankel functions, etc.

The power generated inside the nanoparticle of radius $r_1$ is given by Poynting's theorem as
\begin{equation}
P_1=\displaystyle\frac{1}{2}\omega\epsilon_0\Im\{\epsilon_1\}\int_{V_{r_1}}\left| \vec{E}(\vec{r}) \right|^2\diff v 
=\frac{1}{2}\omega\epsilon_0\Im\{\epsilon_1\}\sum_{l=1}^{\infty}\sum_{m=-l}^{l}\sum_{\tau=1}^2  W_{\tau l}(k_1,r_1)\left| a_{\tau ml}^{(1)}\right|^2,
\end{equation}
where the orthogonality of the regular spherical waves have been used and where $W_{\tau l}(k,r_1)$ is defined in
\eqref{eq:Wtauldef}, \eqref{eq:W1ldef1} and \eqref{eq:W2ldef} in Appendix \ref{sect:spherical}.

It is assumed that there is a small cluster of nanoparticles centered at the origin and where 
the particles have a volume fraction $f_1$. 
It is furthermore assumed that the scattering from the nanocluster is weak so that the multiple scattering effects of the many particles can be neglected as well as
any possible reflections outside the exterior sphere of radius $r_2$.
The number of particles per unit volume inside the cluster is 
\begin{equation}
n=\frac{f_1}{4\pi r_1^3/3},
\end{equation}
and the local heating (in \unit{W/m^3}) is hence given by
\begin{equation}\label{eq:Plocr1def}
P_{\rm loc}(r_1)=
nP_1=\frac{3f_1}{4\pi r_1^3}\frac{1}{2}\omega\epsilon_0\Im\{\epsilon_1\}\sum_{l=1}^{\infty}\sum_{m=-l}^{l}\sum_{\tau=1}^2  W_{\tau l}(k_1,r_1)\left| r_{\tau l} \right|^2\left| a_{\tau ml}^{(2)}\right|^2,
\end{equation}
where we have also employed that $a_{\tau ml}^{(1)}=r_{\tau l}a_{\tau ml}^{(2)}$.

The mean background loss at radius $r$ in medium 2  (in \unit{W/m^3}) is given by
\begin{equation}\label{eq:Pbdef}
P_{\rm b}(r)=\displaystyle \frac{1}{4\pi r^2}\int_{S_r}\frac{1}{2}\omega\epsilon_0\Im\{\epsilon\}\left| \vec{E}(\vec{r}) \right|^2\diff S
=\frac{1}{8\pi}\omega\epsilon_0\Im\{\epsilon\}\int_{\Omega}\left| \vec{E}(\vec{r}) \right|^2\diff \Omega,
\end{equation}
where $S_r$ denotes the spherical boundary of radius $r$ and $\diff S=r^2\diff\Omega$. 
By exploiting the orthogonality of the vector spherical waves \eqref{eq:vorthogonal1}
the mean background loss can now be expressed as
\begin{equation}\label{eq:Pbdef2}
P_{\rm b}(r)=\frac{1}{8\pi}\omega\epsilon_0\Im\{\epsilon\}
\sum_{l=1}^{\infty}\sum_{m=-l}^{l}\sum_{\tau=1}^2  S_{\tau l}(k,r)\left| a_{\tau ml}^{(2)}\right|^2,
\end{equation}
where $S_{\tau l}(k,r)$ is defined in \eqref{eq:Stauldef} in Appendix \ref{sect:spherical}.

The optimal near field is now defined by the maximization of the power ratio $P_{\rm loc}(r_1)/P_{\rm b}(r_2)$. 
This power ratio is a generalized Rayleigh quotient and the problem is hence equivalent to finding the maximum eigenvalue
in the corresponding (diagonal) generalized eigenvalue problem as follows
\begin{equation}\label{eq:Plocr1Pbr2}
\displaystyle \max_{\left|a_{\tau ml}^{(2)}\right|^2} \frac{P_{\rm loc}(r_1)}{P_{\rm b}(r_2)} 
= \frac{3f_1}{r_1^3}\frac{\Im\{\epsilon_1\}}{\Im\{\epsilon\}} \max_{\tau, l} \frac{W_{\tau l}(k_1,r_1)\left| r_{\tau l} \right|^2}{S_{\tau l}(k,r_2)}.
\end{equation}

\section{Numerical examples}

\subsection{Simplified quasistatic analysis and loss-less Mie Theory}
\begin{figure}[htb]
\begin{center}
\input{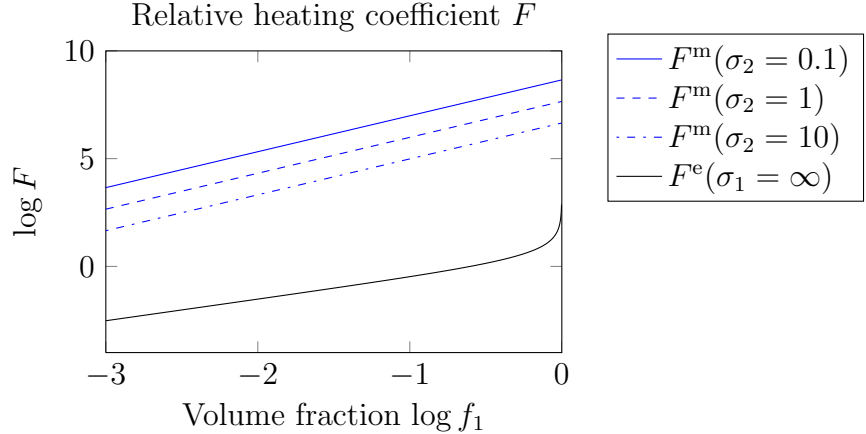}
\end{center}
\vspace{-7mm}
\caption{Relative heating coefficient $F^{\rm m}$ and $F^{\rm e}$ as a function of volume fraction $f_1$ (in log-log-scale).}
\label{fig:matfig1}
\end{figure}

In Figure \ref{fig:matfig1} is shown the relative heating coefficients \eqref{eq:Fedef} and  \eqref{eq:Fmdef} for the quasi-electrostatic and the quasi-magnetostatic cases,
respectively, plotted as functions of the volume fraction $f_1$. 
Here, $F^{\rm m}$ indicates the magnetic case \eqref{eq:Fmdef} with $\sigma_1=4.52\cdot 10^7$\unit{S/m} (gold nanoparticles) 
and $\sigma_2\in\{10^{-1},10^{0},10^{1}\}$\unit{S/m}, and $F^{\rm e}$ indicates the electric case \eqref{eq:Fedef} with $\sigma_1=\infty$.
This result indicates a significant local heating based on magnetic induction, and only negligible electric heating except for the case when the volume fraction $f_1$
is very close to unity.

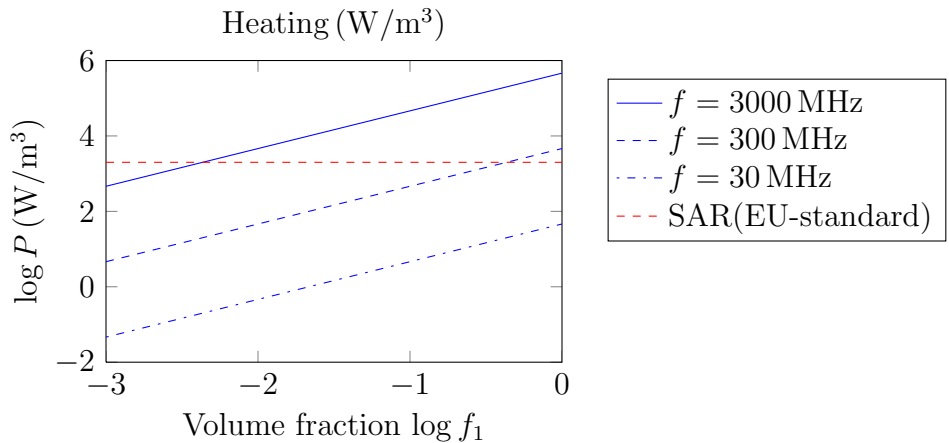
\begin{figure}[htb]
\begin{center}
%
%
\begin{tikzpicture}

\begin{axis}[%
width=6cm,
height=4cm,
scale only axis,
xmin=-3,
xmax=0,
xlabel style={yshift=0cm},
xlabel={Volume fraction $\log f_1$},
ymin=-2,
ymax=6,
ylabel style={yshift=-0.2cm},
ylabel={$\log P$\unit{(W/m^3)}},
title style={yshift=-0.1cm},
title={Heating\unit{(W/m^3)}},
legend style={at={(1.1,0.4)},anchor=south west,draw=black,fill=white,legend cell align=left}
]
\addplot [color=blue,solid]
  table[row sep=crcr]{-3	2.66513316872617\\
-2.95	2.71513316872617\\
-2.9	2.76513316872617\\
-2.85	2.81513316872617\\
-2.8	2.86513316872617\\
-2.75	2.91513316872617\\
-2.7	2.96513316872617\\
-2.65	3.01513316872617\\
-2.6	3.06513316872617\\
-2.55	3.11513316872617\\
-2.5	3.16513316872617\\
-2.45	3.21513316872617\\
-2.4	3.26513316872617\\
-2.35	3.31513316872617\\
-2.3	3.36513316872617\\
-2.25	3.41513316872617\\
-2.2	3.46513316872617\\
-2.15	3.51513316872617\\
-2.1	3.56513316872617\\
-2.05	3.61513316872617\\
-2	3.66513316872617\\
-1.95	3.71513316872617\\
-1.9	3.76513316872617\\
-1.85	3.81513316872617\\
-1.8	3.86513316872617\\
-1.75	3.91513316872617\\
-1.7	3.96513316872617\\
-1.65	4.01513316872617\\
-1.6	4.06513316872617\\
-1.55	4.11513316872617\\
-1.5	4.16513316872617\\
-1.45	4.21513316872617\\
-1.4	4.26513316872617\\
-1.35	4.31513316872617\\
-1.3	4.36513316872617\\
-1.25	4.41513316872617\\
-1.2	4.46513316872617\\
-1.15	4.51513316872617\\
-1.1	4.56513316872617\\
-1.05	4.61513316872617\\
-1	4.66513316872617\\
-0.95	4.71513316872617\\
-0.9	4.76513316872617\\
-0.85	4.81513316872617\\
-0.8	4.86513316872617\\
-0.75	4.91513316872617\\
-0.7	4.96513316872617\\
-0.65	5.01513316872617\\
-0.6	5.06513316872617\\
-0.55	5.11513316872617\\
-0.5	5.16513316872617\\
-0.45	5.21513316872617\\
-0.4	5.26513316872617\\
-0.35	5.31513316872617\\
-0.3	5.36513316872617\\
-0.25	5.41513316872617\\
-0.2	5.46513316872617\\
-0.15	5.51513316872617\\
-0.1	5.56513316872617\\
-0.05	5.61513316872617\\
0	5.66513316872617\\
};
\addlegendentry{$f=3000\unit{MHz}$};

\addplot [color=blue,dashed]
  table[row sep=crcr]{-3	0.665133168726168\\
-2.95	0.715133168726168\\
-2.9	0.765133168726168\\
-2.85	0.815133168726168\\
-2.8	0.865133168726168\\
-2.75	0.915133168726168\\
-2.7	0.965133168726168\\
-2.65	1.01513316872617\\
-2.6	1.06513316872617\\
-2.55	1.11513316872617\\
-2.5	1.16513316872617\\
-2.45	1.21513316872617\\
-2.4	1.26513316872617\\
-2.35	1.31513316872617\\
-2.3	1.36513316872617\\
-2.25	1.41513316872617\\
-2.2	1.46513316872617\\
-2.15	1.51513316872617\\
-2.1	1.56513316872617\\
-2.05	1.61513316872617\\
-2	1.66513316872617\\
-1.95	1.71513316872617\\
-1.9	1.76513316872617\\
-1.85	1.81513316872617\\
-1.8	1.86513316872617\\
-1.75	1.91513316872617\\
-1.7	1.96513316872617\\
-1.65	2.01513316872617\\
-1.6	2.06513316872617\\
-1.55	2.11513316872617\\
-1.5	2.16513316872617\\
-1.45	2.21513316872617\\
-1.4	2.26513316872617\\
-1.35	2.31513316872617\\
-1.3	2.36513316872617\\
-1.25	2.41513316872617\\
-1.2	2.46513316872617\\
-1.15	2.51513316872617\\
-1.1	2.56513316872617\\
-1.05	2.61513316872617\\
-1	2.66513316872617\\
-0.95	2.71513316872617\\
-0.9	2.76513316872617\\
-0.85	2.81513316872617\\
-0.8	2.86513316872617\\
-0.75	2.91513316872617\\
-0.7	2.96513316872617\\
-0.65	3.01513316872617\\
-0.6	3.06513316872617\\
-0.55	3.11513316872617\\
-0.5	3.16513316872617\\
-0.45	3.21513316872617\\
-0.4	3.26513316872617\\
-0.35	3.31513316872617\\
-0.3	3.36513316872617\\
-0.25	3.41513316872617\\
-0.2	3.46513316872617\\
-0.15	3.51513316872617\\
-0.1	3.56513316872617\\
-0.05	3.61513316872617\\
0	3.66513316872617\\
};
\addlegendentry{$f=300\unit{MHz}$};

\addplot [color=blue,dash pattern=on 1pt off 3pt on 3pt off 3pt]
  table[row sep=crcr]{-3	-1.33486683127383\\
-2.95	-1.28486683127383\\
-2.9	-1.23486683127383\\
-2.85	-1.18486683127383\\
-2.8	-1.13486683127383\\
-2.75	-1.08486683127383\\
-2.7	-1.03486683127383\\
-2.65	-0.984866831273832\\
-2.6	-0.934866831273832\\
-2.55	-0.884866831273832\\
-2.5	-0.834866831273832\\
-2.45	-0.784866831273832\\
-2.4	-0.734866831273832\\
-2.35	-0.684866831273832\\
-2.3	-0.634866831273832\\
-2.25	-0.584866831273832\\
-2.2	-0.534866831273832\\
-2.15	-0.484866831273832\\
-2.1	-0.434866831273832\\
-2.05	-0.384866831273832\\
-2	-0.334866831273832\\
-1.95	-0.284866831273832\\
-1.9	-0.234866831273832\\
-1.85	-0.184866831273832\\
-1.8	-0.134866831273832\\
-1.75	-0.0848668312738321\\
-1.7	-0.0348668312738319\\
-1.65	0.015133168726168\\
-1.6	0.0651331687261681\\
-1.55	0.115133168726168\\
-1.5	0.165133168726168\\
-1.45	0.215133168726168\\
-1.4	0.265133168726168\\
-1.35	0.315133168726168\\
-1.3	0.365133168726168\\
-1.25	0.415133168726168\\
-1.2	0.465133168726168\\
-1.15	0.515133168726168\\
-1.1	0.565133168726168\\
-1.05	0.615133168726168\\
-1	0.665133168726168\\
-0.95	0.715133168726168\\
-0.9	0.765133168726168\\
-0.85	0.815133168726168\\
-0.8	0.865133168726168\\
-0.75	0.915133168726168\\
-0.7	0.965133168726168\\
-0.65	1.01513316872617\\
-0.6	1.06513316872617\\
-0.55	1.11513316872617\\
-0.5	1.16513316872617\\
-0.45	1.21513316872617\\
-0.4	1.26513316872617\\
-0.35	1.31513316872617\\
-0.3	1.36513316872617\\
-0.25	1.41513316872617\\
-0.2	1.46513316872617\\
-0.15	1.51513316872617\\
-0.1	1.56513316872617\\
-0.05	1.61513316872617\\
0	1.66513316872617\\
};
\addlegendentry{$f=30\unit{MHz}$};

\addplot [color=red,dashed]
  table[row sep=crcr]{-3	3.30102999566398\\
0	3.30102999566398\\
};
\addlegendentry{SAR(EU-standard)};

\end{axis}
\end{tikzpicture}%
\end{center}
\vspace{-7mm}
\caption{Local heating $P_{\rm loc}$  according to the asymptotic Mie theory  \eqref{eq:Ploc2}, plotted 
as a function of volume fraction $f_1$ (in log-log-scale).}
\label{fig:matfig4}
\end{figure}

\begin{figure}[htb]
\begin{center}
%
%
\begin{tikzpicture}

\begin{axis}[%
width=6cm,
height=4cm,
scale only axis,
xmin=-3,
xmax=0,
xlabel style={yshift=0cm},
xlabel={Volume fraction $\log f_1$},
ymin=0,
ymax=10,
ylabel style={yshift=-0.2cm},
ylabel={$\Delta T/\Delta t$ (\unit{\degree C/h})},
title style={yshift=-0.1cm},
title={Temperature increase $\Delta T/\Delta t$ (\unit{\degree C/h})},
legend style={at={(1.1,0.4)},anchor=south west,draw=black,fill=white,legend cell align=left}
]
\addplot [color=blue,solid]
  table[row sep=crcr]{-3	0.398440336631014\\
-2.95	0.447057410638284\\
-2.9	0.501606664868606\\
-2.85	0.562811934783435\\
-2.8	0.631485377128408\\
-2.75	0.708538246759909\\
-2.7	0.794992988443376\\
-2.65	0.891996804074135\\
-2.6	1.00083687534955\\
-2.55	1.12295744388811\\
-2.5	1.25997897543822\\
-2.45	1.41371966247417\\
-2.4	1.58621955050556\\
-2.35	1.7797676082418\\
-2.3	1.99693210081617\\
-2.25	2.24059466910373\\
-2.2	2.51398856734499\\
-2.15	2.82074156646523\\
-2.1	3.16492409239061\\
-2.05	3.55110323812716\\
-2	3.98440336631014\\
-1.95	4.47057410638284\\
-1.9	5.01606664868605\\
-1.85	5.62811934783436\\
-1.8	6.31485377128408\\
-1.75	7.08538246759909\\
-1.7	7.94992988443376\\
-1.65	8.91996804074135\\
-1.6	10.0083687534955\\
-1.55	11.2295744388811\\
-1.5	12.5997897543822\\
-1.45	14.1371966247417\\
-1.4	15.8621955050556\\
-1.35	17.797676082418\\
-1.3	19.9693210081617\\
-1.25	22.4059466910373\\
-1.2	25.1398856734499\\
-1.15	28.2074156646523\\
-1.1	31.6492409239061\\
-1.05	35.5110323812716\\
-1	39.8440336631014\\
-0.95	44.7057410638284\\
-0.9	50.1606664868605\\
-0.85	56.2811934783435\\
-0.8	63.1485377128408\\
-0.75	70.8538246759909\\
-0.7	79.4992988443376\\
-0.65	89.1996804074135\\
-0.6	100.083687534955\\
-0.55	112.295744388811\\
-0.5	125.997897543822\\
-0.45	141.371966247417\\
-0.4	158.621955050556\\
-0.35	177.97676082418\\
-0.3	199.693210081617\\
-0.25	224.059466910373\\
-0.2	251.398856734499\\
-0.15	282.074156646523\\
-0.1	316.492409239061\\
-0.05	355.110323812716\\
0	398.440336631014\\
};
\addlegendentry{$f=3000\unit{MHz}$};

\addplot [color=blue,dashed]
  table[row sep=crcr]{-3	0.00398440336631014\\
-2.95	0.00447057410638284\\
-2.9	0.00501606664868606\\
-2.85	0.00562811934783436\\
-2.8	0.00631485377128408\\
-2.75	0.00708538246759909\\
-2.7	0.00794992988443376\\
-2.65	0.00891996804074136\\
-2.6	0.0100083687534955\\
-2.55	0.0112295744388811\\
-2.5	0.0125997897543822\\
-2.45	0.0141371966247417\\
-2.4	0.0158621955050556\\
-2.35	0.017797676082418\\
-2.3	0.0199693210081617\\
-2.25	0.0224059466910373\\
-2.2	0.0251398856734499\\
-2.15	0.0282074156646523\\
-2.1	0.0316492409239061\\
-2.05	0.0355110323812716\\
-2	0.0398440336631014\\
-1.95	0.0447057410638284\\
-1.9	0.0501606664868606\\
-1.85	0.0562811934783436\\
-1.8	0.0631485377128408\\
-1.75	0.0708538246759909\\
-1.7	0.0794992988443376\\
-1.65	0.0891996804074136\\
-1.6	0.100083687534955\\
-1.55	0.112295744388811\\
-1.5	0.125997897543822\\
-1.45	0.141371966247417\\
-1.4	0.158621955050556\\
-1.35	0.17797676082418\\
-1.3	0.199693210081617\\
-1.25	0.224059466910373\\
-1.2	0.251398856734499\\
-1.15	0.282074156646523\\
-1.1	0.316492409239061\\
-1.05	0.355110323812716\\
-1	0.398440336631014\\
-0.95	0.447057410638284\\
-0.9	0.501606664868605\\
-0.85	0.562811934783436\\
-0.8	0.631485377128408\\
-0.75	0.708538246759909\\
-0.7	0.794992988443376\\
-0.65	0.891996804074135\\
-0.6	1.00083687534955\\
-0.55	1.12295744388811\\
-0.5	1.25997897543822\\
-0.45	1.41371966247417\\
-0.4	1.58621955050556\\
-0.35	1.7797676082418\\
-0.3	1.99693210081617\\
-0.25	2.24059466910373\\
-0.2	2.51398856734499\\
-0.15	2.82074156646523\\
-0.1	3.16492409239061\\
-0.05	3.55110323812716\\
0	3.98440336631014\\
};
\addlegendentry{$f=300\unit{MHz}$};

\addplot [color=blue,dash pattern=on 1pt off 3pt on 3pt off 3pt]
  table[row sep=crcr]{-3	3.98440336631014e-05\\
-2.95	4.47057410638284e-05\\
-2.9	5.01606664868606e-05\\
-2.85	5.62811934783435e-05\\
-2.8	6.31485377128408e-05\\
-2.75	7.08538246759909e-05\\
-2.7	7.94992988443376e-05\\
-2.65	8.91996804074135e-05\\
-2.6	0.000100083687534955\\
-2.55	0.000112295744388811\\
-2.5	0.000125997897543822\\
-2.45	0.000141371966247417\\
-2.4	0.000158621955050556\\
-2.35	0.00017797676082418\\
-2.3	0.000199693210081617\\
-2.25	0.000224059466910373\\
-2.2	0.000251398856734499\\
-2.15	0.000282074156646523\\
-2.1	0.000316492409239062\\
-2.05	0.000355110323812716\\
-2	0.000398440336631014\\
-1.95	0.000447057410638284\\
-1.9	0.000501606664868606\\
-1.85	0.000562811934783436\\
-1.8	0.000631485377128408\\
-1.75	0.000708538246759909\\
-1.7	0.000794992988443376\\
-1.65	0.000891996804074135\\
-1.6	0.00100083687534955\\
-1.55	0.00112295744388811\\
-1.5	0.00125997897543822\\
-1.45	0.00141371966247417\\
-1.4	0.00158621955050556\\
-1.35	0.0017797676082418\\
-1.3	0.00199693210081617\\
-1.25	0.00224059466910373\\
-1.2	0.00251398856734499\\
-1.15	0.00282074156646523\\
-1.1	0.00316492409239062\\
-1.05	0.00355110323812716\\
-1	0.00398440336631014\\
-0.95	0.00447057410638284\\
-0.9	0.00501606664868605\\
-0.85	0.00562811934783436\\
-0.8	0.00631485377128408\\
-0.75	0.00708538246759909\\
-0.7	0.00794992988443376\\
-0.65	0.00891996804074136\\
-0.6	0.0100083687534955\\
-0.55	0.0112295744388811\\
-0.5	0.0125997897543822\\
-0.45	0.0141371966247417\\
-0.4	0.0158621955050556\\
-0.35	0.017797676082418\\
-0.3	0.0199693210081617\\
-0.25	0.0224059466910373\\
-0.2	0.0251398856734499\\
-0.15	0.0282074156646523\\
-0.1	0.0316492409239062\\
-0.05	0.0355110323812716\\
0	0.0398440336631014\\
};
\addlegendentry{$f=30\unit{MHz}$};

\addplot [color=red,dashed]
  table[row sep=crcr]{-3	1.72290021536253\\
0	1.72290021536253\\
};
\addlegendentry{SAR(EU-standard)};

\end{axis}
\end{tikzpicture}%
\end{center}
\vspace{-7mm}
\caption{Local heating $P_{\rm loc}$ according to the asymptotic Mie theory \eqref{eq:Ploc2}. Same result as in Figure \ref{fig:matfig4}
recalculated as an equivalent temperature increase in water per unit time $\Delta T/\Delta t=P_{\rm loc}3600/c\rho$ (\unit{\degree C/h}) and plotted 
as a function of volume fraction $f_1$ (in log-lin-scale).
Here, $c=4179$\unit{J/kg \degree C } is the specific heat capacity of water and $\rho=1000$\unit{kg/m^3} the specific weight of water.}
\label{fig:matfig41}
\end{figure}
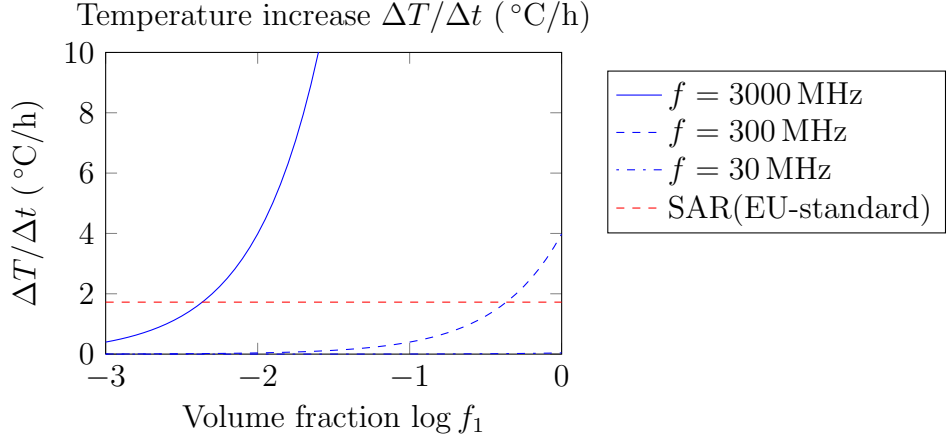

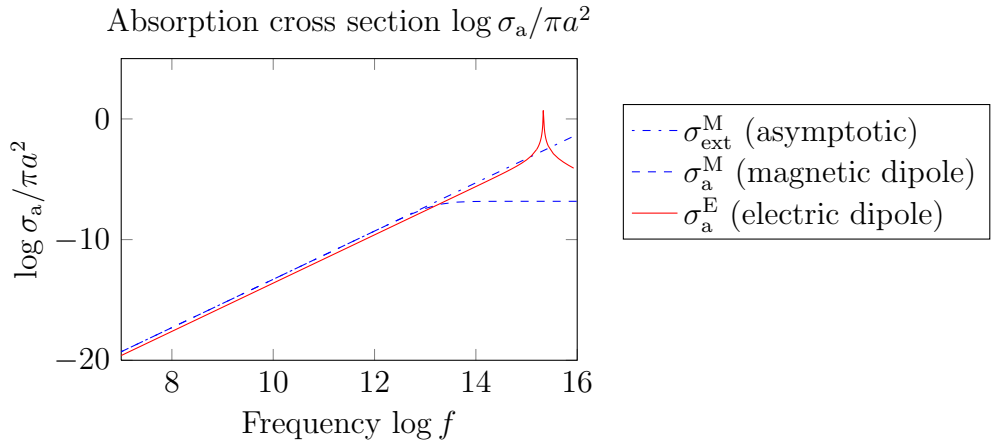
\begin{figure}[htb]
\begin{center}
%
%
\begin{tikzpicture}

\begin{axis}[%
width=6cm,
height=4cm,
scale only axis,
xmin=7,
xmax=16,
xlabel style={yshift=0cm},
xlabel={Frequency $\log f$},
ymin=-20,
ymax=5,
title style={yshift=-0.1cm},
ylabel style={yshift=0cm},
ylabel={$\log \sigma_{\rm a}/\pi a^2$},
title={Absorption cross section $\log \sigma_{\rm a}/\pi a^2$},
legend style={at={(1.1,0.4)},anchor=south west,draw=black,fill=white,legend cell align=left}
]
\addplot [color=blue,dash pattern=on 1pt off 3pt on 3pt off 3pt]
  table[row sep=crcr]{7	-19.2919039697941\\
7.3	-18.6919039697941\\
7.6	-18.0919039697941\\
7.9	-17.4919039697941\\
8.2	-16.8919039697941\\
8.5	-16.2919039697941\\
8.8	-15.6919039697941\\
9.1	-15.0919039697941\\
9.4	-14.4919039697941\\
9.7	-13.8919039697941\\
10	-13.2919039697941\\
10.3	-12.6919039697941\\
10.6	-12.0919039697941\\
10.9	-11.4919039697941\\
11.2	-10.8919039697941\\
11.5	-10.2919039697941\\
11.8	-9.69190396979409\\
12.1	-9.09190396979409\\
12.4	-8.49190396979409\\
12.7	-7.89190396979409\\
13	-7.29190396979409\\
13.3	-6.69190396979409\\
13.6	-6.09190396979409\\
13.9	-5.49190396979409\\
14.2	-4.89190396979409\\
14.5	-4.29190396979409\\
14.8	-3.69190396979409\\
15.1	-3.09190396979409\\
15.4	-2.49190396979409\\
15.7	-1.89190396979409\\
16	-1.29190396979409\\
};
\addlegendentry{$\sigma_{\rm ext}^{\rm M}$ (asymptotic)};

\addplot [color=blue,dashed]
  table[row sep=crcr]{7	-19.291904078023\\
7.3	-18.6919039225507\\
7.6	-18.0919039239969\\
7.9	-17.4919039581582\\
8.2	-16.8919039636313\\
8.5	-16.2919039711445\\
8.8	-15.6919039750164\\
9.1	-15.0919039703686\\
9.4	-14.4919039792786\\
9.7	-13.8919040070641\\
10	-13.291904118194\\
10.3	-12.6919045605938\\
10.6	-12.0919063217931\\
10.9	-11.4919133330737\\
11.2	-10.8919412445435\\
11.5	-10.2920523442207\\
11.8	-9.69249435846459\\
12.1	-9.09424960190179\\
12.4	-8.50116785288933\\
12.7	-7.92766658222924\\
13	-7.41956039548611\\
13.3	-7.06487937840556\\
13.6	-6.89914324728237\\
13.9	-6.84524332515512\\
14.2	-6.83058389719148\\
14.5	-6.82682374398457\\
14.8	-6.82587897456011\\
15.1	-6.82566076033162\\
15.4	-6.8256832716374\\
15.7	-6.82599684957401\\
16	-6.82730163057019\\
};
\addlegendentry{$\sigma_{\rm a}^{\rm M}$ (magnetic dipole)};

\addplot [color=red,solid]
  table[row sep=crcr]{7	-19.6061829294367\\
7.1	-19.4061797445208\\
7.2	-19.2061799722235\\
7.3	-19.0061798670079\\
7.4	-18.8061804945953\\
7.5	-18.6061792282069\\
7.6	-18.4061801248421\\
7.7	-18.2061796663658\\
7.8	-18.0061793214473\\
7.9	-17.8061793628745\\
8	-17.6061795618634\\
8.1	-17.406179743144\\
8.2	-17.2061795110964\\
8.3	-17.0061794502345\\
8.4	-16.8061795283246\\
8.5	-16.6061794182577\\
8.6	-16.4061794764583\\
8.7	-16.2061794694061\\
8.8	-16.0061794009312\\
8.9	-15.8061794099043\\
9	-15.6061794389921\\
9.1	-15.4061794502805\\
9.2	-15.2061794482379\\
9.3	-15.006179445549\\
9.4	-14.8061794537148\\
9.5	-14.6061794368512\\
9.6	-14.4061794420529\\
9.7	-14.2061794463673\\
9.8	-14.0061794426283\\
9.9	-13.8061794365109\\
10	-13.6061794381217\\
10.1	-13.4061794369974\\
10.2	-13.2061794393169\\
10.3	-13.0061794372964\\
10.4	-12.8061794382438\\
10.5	-12.6061794388965\\
10.6	-12.4061794379165\\
10.7	-12.2061794377357\\
10.8	-12.0061794373449\\
10.9	-11.8061794366965\\
11	-11.6061794360648\\
11.1	-11.4061794348182\\
11.2	-11.2061794331719\\
11.3	-11.0061794304246\\
11.4	-10.8061794259956\\
11.5	-10.6061794191763\\
11.6	-10.4061794081312\\
11.7	-10.206179390759\\
11.8	-10.0061793632052\\
11.9	-9.80617931946284\\
12	-9.60617925022428\\
12.1	-9.40617914050058\\
12.2	-9.2061789665601\\
12.3	-9.00617869088758\\
12.4	-8.80617825395435\\
12.5	-8.60617756151065\\
12.6	-8.40617646405154\\
12.7	-8.20617472472995\\
12.8	-8.00617196814116\\
12.9	-7.80616759934993\\
13	-7.60616067551206\\
13.1	-7.40614970234871\\
13.2	-7.20613231157769\\
13.3	-7.00610474950176\\
13.4	-6.80606106617546\\
13.5	-6.60599182976208\\
13.6	-6.40588208751448\\
13.7	-6.20570813050549\\
13.8	-6.00543235669584\\
13.9	-5.80499510581688\\
14	-5.60430165901926\\
14.1	-5.40320148468243\\
14.2	-5.20145496601622\\
14.3	-4.99867970545119\\
14.4	-4.79426296560596\\
14.5	-4.58721656350145\\
14.6	-4.37593031589636\\
14.7	-4.15773682840918\\
14.8	-3.92809786610783\\
14.9	-3.6789475235911\\
15	-3.39486056325024\\
15.1	-3.04227506446461\\
15.2	-2.52703077071933\\
};
\addlegendentry{$\sigma_{\rm a}^{\rm E}$ (electric dipole)};

\addplot [color=red,solid,forget plot]
  table[row sep=crcr]{15.2	-2.52703077071933\\
15.21	-2.45708933038746\\
15.22	-2.38152221141665\\
15.23	-2.29925652783216\\
15.24	-2.20888945482984\\
15.25	-2.10853747721262\\
15.26	-1.99558834246575\\
15.27	-1.86626910049624\\
15.28	-1.71483781629639\\
15.29	-1.53192147720797\\
15.3	-1.30062334399204\\
15.31	-0.985461005782908\\
15.32	-0.488742291521765\\
15.33	0.700738027085951\\
15.34	-0.00786058407943163\\
15.35	-0.74354603829842\\
15.36	-1.13910809489677\\
15.37	-1.41062668994177\\
15.38	-1.61765736668268\\
15.39	-1.7851438219868\\
15.4	-1.92591140095237\\
15.41	-2.04742912265512\\
15.42	-2.15442973441907\\
15.43	-2.25010149318561\\
};
\addplot [color=red,solid,forget plot]
  table[row sep=crcr]{15.43	-2.25010149318561\\
15.53	-2.88692902210982\\
15.63	-3.28046091146176\\
15.73	-3.5837437637567\\
15.83	-3.84338491166238\\
15.93	-4.07943048129953\\
};
\end{axis}
\end{tikzpicture}%
\end{center}
\vspace{-7mm}
\caption{Absorption cross section $\sigma_{\rm a}$ of a gold nanoparticle plotted as a function of frequency (in log-log-scale). 
Here, the radius of the particle is $a=0.8$\unit{nm}, the absorption cross section is normalized with $\pi a^2$
and a Drude model is used to model the material dispersion of gold.}
\label{fig:matfig20}
\end{figure}

To study the potential of inductive heating we consider also the Mie theory as outlined in section \ref{sect:Mie} above where
the result \eqref{eq:Ploc2} agrees very well with the quasi-magnetostatic theory \eqref{eq:Ploc}.
In Figure \ref{fig:matfig4} is shown the local heating $P_{\rm loc}$ according to the asymptotic Mie theory \eqref{eq:Ploc2}, plotted as a function of volume fraction $f_1$.
Here, the gold nanoparticles are placed in vacuum (or air), $r_1=0.8$\unit{nm}, $\sigma_1=4.52\cdot 10^7$\unit{S/m}, $B_0=0.03$\unit{T} and $f\in\{30,300,3000\}$\unit{MHz}.
The European SAR limit of 2000\unit{W/m^3} is also included in the plot. 
In Figure \ref{fig:matfig41} is shown the same results recalculated as an equivalent temperature 
increase in water per unit time (\unit{\degree C/h}). Again, the results indicate a significant local heating based on magnetic induction
given that a sufficiently high magnetic flux density can be employed at sufficiently high frequency.

As an illustration of the Mie theory for higher frequencies we consider the following Drude model for gold
\begin{equation}
\epsilon(\omega)=1+\iu\frac{\sigma_1}{\omega\epsilon_0}\frac{1}{1-\iu\omega\tau}
\end{equation}
where $\sigma_1=4.52 \cdot 10^7$\unit{S/m} is the static conductivity and
$\tau=9.3 \cdot 10^{-15}$\unit{s} is the mean collision time for electrons in gold \cite{Johnson+Christy1972}.
In Figure \ref{fig:matfig20} is shown the magnetic and electric absorption cross sections
$\sigma_{\rm a}^{\rm M}$ and $\sigma_{\rm a}^{\rm E}$ given by the exact Mie
theory \eqref{eq:signmas} and \eqref{eq:signmaext} respectively, and where $\sigma_{\rm a}=\sigma_{\rm ext}-\sigma_{\rm s}$.
These expressions converged well  already with $l=1$ in this example.
The plot also shows the asymptotic expression \eqref{eq:sigmaextM} for $\sigma_{\rm ext}^{\rm M}$ based on the static conductivity of gold,
and which apparently is valid up to about $10^{13}$\unit{Hz}. It is noted that the cut-off frequency in
$\sigma_{\rm a}^{\rm M}$ at about $10^{13}$\unit{Hz} corresponds approximately to the collision frequency $1/2\pi \tau$ of the Drude model.
It is also noted that there is a plasmonic resonance slightly above $10^{15}$\unit{Hz} (in the ultraviolet region)
which is due to the Drude model.


\subsection{Optimal near field and skin effect}

The following numerical example is employed to evaluate the physical possibility of using electromagnetic waves in the radio frequency spectrum to
heat gold nanoparticles. The nanoparticles are assumed to be immersed in a lossy medium and hence subjected to the skin effect. 
The frequency is chosen to $f=13.56$ \unit{MHz} due to common regulations.
The gold nanoparticle is modeled with a complex valued relative permittivity \eqref{eq:epsilon1def}
with $\sigma_1=4.52 \cdot 10^7$\unit{S/m}. The exterior lossy medium is modelled as salty water with a complex valued relative permittivity \eqref{eq:epsilondef}
with $\epsilon_{\infty}=5.27$, $\epsilon_{\rm s}=80$, $\tau=1 \cdot 10^{-11}$\unit{s} and $\sigma\in\{1,10,100\}$\unit{S/m}.


%
%
%
%
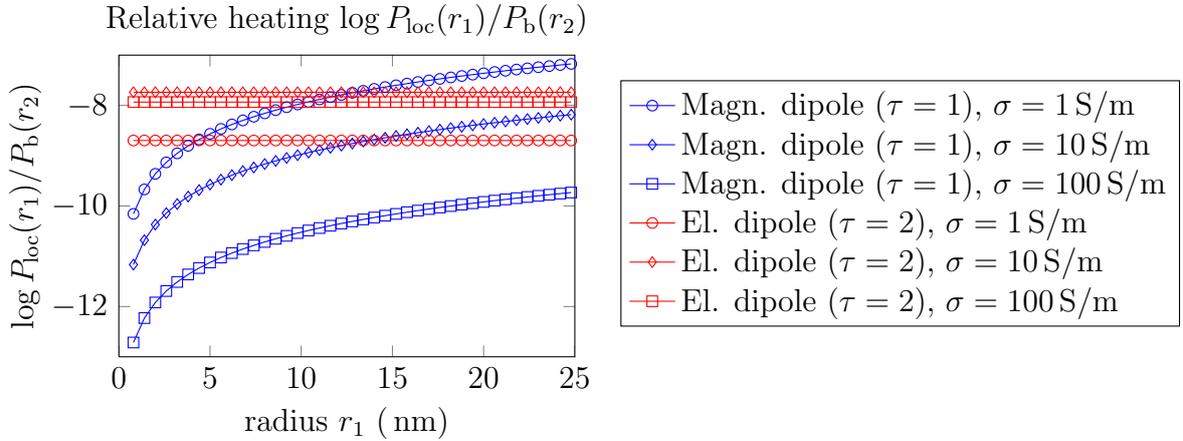
\begin{figure}[htb]
\begin{center}
%
%
\begin{tikzpicture}

\begin{axis}[%
width=6cm,
height=4cm,
scale only axis,
xmin=0,
xmax=25,
xlabel style={yshift=0cm},
xlabel={radius $r_1$ (\unit{nm})},
ymin=-13,
ymax=-7,
ylabel style={yshift=0cm},
ylabel={$\log P_{\rm loc}(r_1)/P_{\rm b}(r_2)$},
title style={yshift=-0.1cm},
title={Relative heating $\log P_{\rm loc}(r_1)/P_{\rm b}(r_2)$},
legend style={at={(1.1,0.1)},anchor=south west,draw=black,fill=white,legend cell align=left}
]
\addplot [color=blue,solid,mark=o,mark options={solid}]
  table[row sep=crcr]{0.8	-10.1571768351773\\
1.4	-9.67110073780476\\
2	-9.36129681783327\\
2.6	-9.1334101132196\\
3.2	-8.95305685252142\\
3.8	-8.80378961592762\\
4.4	-8.67645145618886\\
5	-8.5654168004892\\
5.6	-8.46698075514883\\
6.2	-8.37857343016473\\
6.8	-8.29833898374876\\
7.4	-8.22489336969928\\
8	-8.15717683517735\\
8.6	-8.0943599066741\\
9.2	-8.03578115447013\\
9.8	-7.98090465777624\\
10.4	-7.92929013056368\\
11	-7.88057143884479\\
11.6	-7.8344408307074\\
12.2	-7.79063714781174\\
12.8	-7.7489368698655\\
13.4	-7.70914721243162\\
14	-7.67110073780476\\
14.6	-7.63465109759237\\
15.2	-7.5996696332717\\
15.8	-7.5660426352524\\
16.4	-7.53366911306585\\
17	-7.5024589664047\\
17.6	-7.47233147353294\\
18.2	-7.4432140331911\\
18.8	-7.41504111063389\\
19.4	-7.3877533493008\\
20	-7.36129681783329\\
20.6	-7.33562236842294\\
21.2	-7.31068508730375\\
21.8	-7.28644382195205\\
22.4	-7.26286077249293\\
23	-7.23990113712608\\
23.6	-7.21753280322105\\
24.2	-7.1957260772004\\
24.8	-7.17445344750884\\
};
\addlegendentry{Magn. dipole ($\tau=1$), $\sigma=1$\unit{S/m}};

\addplot [color=blue,solid,mark=diamond,mark options={solid}]
  table[row sep=crcr]{0.8	-11.1659043163146\\
1.4	-10.679828218942\\
2	-10.3700242989705\\
2.6	-10.1421375943569\\
3.2	-9.96178433365868\\
3.8	-9.81251709706487\\
4.4	-9.68517893732612\\
5	-9.57414428162646\\
5.6	-9.47570823628609\\
6.2	-9.38730091130199\\
6.8	-9.30706646488602\\
7.4	-9.23362085083654\\
8	-9.16590431631461\\
8.6	-9.10308738781136\\
9.2	-9.04450863560738\\
9.8	-8.9896321389135\\
10.4	-8.93801761170093\\
11	-8.88929891998204\\
11.6	-8.84316831184466\\
12.2	-8.799364628949\\
12.8	-8.75766435100276\\
13.4	-8.71787469356888\\
14	-8.67982821894202\\
14.6	-8.64337857872962\\
15.2	-8.60839711440895\\
15.8	-8.57477011638965\\
16.4	-8.5423965942031\\
17	-8.51118644754195\\
17.6	-8.4810589546702\\
18.2	-8.45194151432835\\
18.8	-8.42376859177115\\
19.4	-8.39648083043805\\
20	-8.37002429897055\\
20.6	-8.3443498495602\\
21.2	-8.31941256844101\\
21.8	-8.2951713030893\\
22.4	-8.27158825363019\\
23	-8.24862861826333\\
23.6	-8.22626028435831\\
24.2	-8.20445355833766\\
24.8	-8.1831809286461\\
};
\addlegendentry{Magn. dipole ($\tau=1$), $\sigma=10$\unit{S/m}};

\addplot [color=blue,solid,mark=square,mark options={solid}]
  table[row sep=crcr]{0.8	-12.7154172333114\\
1.4	-12.2293411359388\\
2	-11.9195372159673\\
2.6	-11.6916505113536\\
3.2	-11.5112972506555\\
3.8	-11.3620300140616\\
4.4	-11.2346918543229\\
5	-11.1236571986232\\
5.6	-11.0252211532829\\
6.2	-10.9368138282988\\
6.8	-10.8565793818828\\
7.4	-10.7831337678333\\
8	-10.7154172333114\\
8.6	-10.6526003048081\\
9.2	-10.5940215526042\\
9.8	-10.5391450559103\\
10.4	-10.4875305286977\\
11	-10.4388118369788\\
11.6	-10.3926812288414\\
12.2	-10.3488775459458\\
12.8	-10.3071772679995\\
13.4	-10.2673876105657\\
14	-10.2293411359388\\
14.6	-10.1928914957264\\
15.2	-10.1579100314057\\
15.8	-10.1242830333864\\
16.4	-10.0919095111999\\
17	-10.0606993645387\\
17.6	-10.030571871667\\
18.2	-10.0014544313251\\
18.8	-9.97328150876792\\
19.4	-9.94599374743483\\
20	-9.91953721596732\\
20.6	-9.89386276655698\\
21.2	-9.86892548543779\\
21.8	-9.84468422008608\\
22.4	-9.82110117062697\\
23	-9.79814153526011\\
23.6	-9.77577320135508\\
24.2	-9.75396647533444\\
24.8	-9.73269384564287\\
};
\addlegendentry{Magn. dipole ($\tau=1$), $\sigma=100$\unit{S/m}};

\addplot [color=red,solid,mark=o,mark options={solid}]
  table[row sep=crcr]{0.8	-8.69737290578941\\
1.4	-8.69737290578941\\
2	-8.69737290578941\\
2.6	-8.69737290578941\\
3.2	-8.69737290578941\\
3.8	-8.69737290578941\\
4.4	-8.69737290578941\\
5	-8.69737290578941\\
5.6	-8.69737290578941\\
6.2	-8.69737290578941\\
6.8	-8.69737290578941\\
7.4	-8.69737290578941\\
8	-8.69737290578941\\
8.6	-8.69737290578941\\
9.2	-8.69737290578941\\
9.8	-8.69737290578941\\
10.4	-8.69737290578941\\
11	-8.69737290578941\\
11.6	-8.69737290578941\\
12.2	-8.69737290578941\\
12.8	-8.69737290578941\\
13.4	-8.69737290578941\\
14	-8.69737290578941\\
14.6	-8.69737290578941\\
15.2	-8.69737290578941\\
15.8	-8.69737290578941\\
16.4	-8.69737290578941\\
17	-8.69737290578941\\
17.6	-8.6973729057894\\
18.2	-8.69737290578941\\
18.8	-8.69737290578941\\
19.4	-8.6973729057894\\
20	-8.69737290578941\\
20.6	-8.6973729057894\\
21.2	-8.6973729057894\\
21.8	-8.6973729057894\\
22.4	-8.6973729057894\\
23	-8.6973729057894\\
23.6	-8.6973729057894\\
24.2	-8.6973729057894\\
24.8	-8.6973729057894\\
};
\addlegendentry{El. dipole ($\tau=2$), $\sigma=1$\unit{S/m}};

\addplot [color=red,solid,mark=diamond,mark options={solid}]
  table[row sep=crcr]{0.8	-7.7385101787281\\
1.4	-7.7385101787281\\
2	-7.7385101787281\\
2.6	-7.7385101787281\\
3.2	-7.7385101787281\\
3.8	-7.7385101787281\\
4.4	-7.7385101787281\\
5	-7.7385101787281\\
5.6	-7.7385101787281\\
6.2	-7.73851017872811\\
6.8	-7.7385101787281\\
7.4	-7.7385101787281\\
8	-7.7385101787281\\
8.6	-7.7385101787281\\
9.2	-7.7385101787281\\
9.8	-7.7385101787281\\
10.4	-7.7385101787281\\
11	-7.7385101787281\\
11.6	-7.7385101787281\\
12.2	-7.7385101787281\\
12.8	-7.7385101787281\\
13.4	-7.7385101787281\\
14	-7.7385101787281\\
14.6	-7.7385101787281\\
15.2	-7.7385101787281\\
15.8	-7.7385101787281\\
16.4	-7.7385101787281\\
17	-7.7385101787281\\
17.6	-7.7385101787281\\
18.2	-7.7385101787281\\
18.8	-7.7385101787281\\
19.4	-7.7385101787281\\
20	-7.7385101787281\\
20.6	-7.7385101787281\\
21.2	-7.7385101787281\\
21.8	-7.7385101787281\\
22.4	-7.7385101787281\\
23	-7.73851017872809\\
23.6	-7.73851017872809\\
24.2	-7.73851017872809\\
24.8	-7.73851017872809\\
};
\addlegendentry{El. dipole ($\tau=2$), $\sigma=10$\unit{S/m}};

\addplot [color=red,solid,mark=square,mark options={solid}]
  table[row sep=crcr]{0.8	-7.93350085640026\\
1.4	-7.93350085640026\\
2	-7.93350085640026\\
2.6	-7.93350085640026\\
3.2	-7.93350085640026\\
3.8	-7.93350085640026\\
4.4	-7.93350085640026\\
5	-7.93350085640026\\
5.6	-7.93350085640026\\
6.2	-7.93350085640026\\
6.8	-7.93350085640026\\
7.4	-7.93350085640026\\
8	-7.93350085640026\\
8.6	-7.93350085640026\\
9.2	-7.93350085640026\\
9.8	-7.93350085640026\\
10.4	-7.93350085640026\\
11	-7.93350085640026\\
11.6	-7.93350085640026\\
12.2	-7.93350085640026\\
12.8	-7.93350085640026\\
13.4	-7.93350085640026\\
14	-7.93350085640026\\
14.6	-7.93350085640026\\
15.2	-7.93350085640026\\
15.8	-7.93350085640026\\
16.4	-7.93350085640026\\
17	-7.93350085640026\\
17.6	-7.93350085640026\\
18.2	-7.93350085640026\\
18.8	-7.93350085640026\\
19.4	-7.93350085640025\\
20	-7.93350085640025\\
20.6	-7.93350085640025\\
21.2	-7.93350085640025\\
21.8	-7.93350085640025\\
22.4	-7.93350085640025\\
23	-7.93350085640025\\
23.6	-7.93350085640025\\
24.2	-7.93350085640025\\
24.8	-7.93350085640025\\
};
\addlegendentry{El. dipole ($\tau=2$), $\sigma=100$\unit{S/m}};

\end{axis}
\end{tikzpicture}%
\end{center}
\vspace{-7mm}
\caption{Relative heating $\log P_{\rm loc}(r_1)/P_{\rm b}(r_2)$ for gold nanoparticles with radius $r_1$ ranging from $0.8$\unit{nm} to
$25$\unit{nm} with a volume fraction of $f_1=0.01$ and with exterior design radius $r_2=5$\unit{cm}. Here, $l=1$ and the magnetic and electric dipole fields  are indicated with $\tau=1$
and $\tau=2$, respectively.}
\label{fig:matfig51}
\end{figure}

\begin{figure}[htb]
\begin{center}
%
%
\begin{tikzpicture}

\begin{axis}[%
width=6cm,
height=4cm,
scale only axis,
xmin=0,
xmax=5,
xlabel style={yshift=0cm},
xlabel={radius $r$ (\unit{cm})},
ymin=0,
ymax=1.4,
ylabel style={yshift=0cm},
ylabel={$S_{\tau 1}(k,r)/S_{\tau 1}(k,r_2)$},
title style={yshift=-0.1cm},
title={Skin-effect $S_{\tau 1}(k,r)/S_{\tau 1}(k,r_2)$},
legend style={at={(1.1,0.1)},anchor=south west,draw=black,fill=white,legend cell align=left}
]
\addplot [color=blue,solid,mark=o,mark options={solid}]
  table[row sep=crcr]{0.05	0.000100303071702146\\
0.2	0.00160484137089481\\
0.35	0.00491477432156329\\
0.5	0.0100299865706949\\
0.65	0.0169503000881356\\
0.8	0.0256754744785881\\
0.95	0.0362052074036853\\
1.1	0.048539135114101\\
1.25	0.062676833091651\\
1.4	0.0786178168013278\\
1.55	0.0963615425532008\\
1.7	0.115907408474107\\
1.85	0.13725475558905\\
2	0.160402869012203\\
2.15	0.185350979247422\\
2.3	0.212098263598154\\
2.45	0.240643847686607\\
2.6	0.270986807082073\\
2.75	0.303126169038238\\
2.9	0.337060914339357\\
3.05	0.372789979255114\\
3.2	0.410312257604028\\
3.35	0.449626602925205\\
3.5	0.490731830758279\\
3.65	0.53362672103135\\
3.8	0.578310020556716\\
3.95	0.624780445634198\\
4.1	0.673036684761881\\
4.25	0.723077401453999\\
4.4	0.774901237165809\\
4.55	0.828506814325173\\
4.7	0.883892739470662\\
4.85	0.94105760649591\\
5	1\\
};
\addlegendentry{Magn. dipole ($\tau=1$), $\sigma=1$\unit{S/m}};

\addplot [color=blue,solid,mark=diamond,mark options={solid}]
  table[row sep=crcr]{0.05	9.83032732363333e-05\\
0.2	0.00157284483176176\\
0.35	0.00481678805854361\\
0.5	0.00983003303505259\\
0.65	0.0166124486207726\\
0.8	0.0251639023812761\\
0.95	0.0354843010757174\\
1.1	0.0475736417012539\\
1.25	0.0614320730920312\\
1.4	0.0770599680710355\\
1.55	0.094458006154435\\
1.7	0.113627266810108\\
1.85	0.134569333275011\\
2	0.157286406939962\\
2.15	0.181781432315463\\
2.3	0.208058232598407\\
2.45	0.23612165586707\\
2.6	0.265977731940774\\
2.75	0.297633839951147\\
2.9	0.331098886684042\\
3.05	0.366383495765186\\
3.2	0.403500207778396\\
3.35	0.442463691423064\\
3.5	0.483290965837527\\
3.65	0.526001634237104\\
3.8	0.570618129040101\\
3.95	0.617165968682047\\
4.1	0.665674026348003\\
4.25	0.716174810885073\\
4.4	0.768704760192363\\
4.55	0.823304547423751\\
4.7	0.88001940038004\\
4.85	0.938899434511543\\
5	1\\
};
\addlegendentry{Magn. dipole ($\tau=1$), $\sigma=10$\unit{S/m}};

\addplot [color=blue,solid,mark=square,mark options={solid}]
  table[row sep=crcr]{0.05	2.77365980790903e-05\\
0.2	0.000443785735113376\\
0.35	0.00135913899839534\\
0.5	0.00277413880382588\\
0.65	0.00468997115483162\\
0.8	0.00710951062019971\\
0.95	0.0100384652253762\\
1.1	0.0134868244589332\\
1.25	0.0174706166311898\\
1.4	0.0220139863882874\\
1.55	0.0271516096126696\\
1.7	0.0329314715748719\\
1.85	0.0394180454099781\\
2	0.0466959221780111\\
2.15	0.0548739613835629\\
2.3	0.0640900523974499\\
2.45	0.0745166033556489\\
2.6	0.0863669055428653\\
2.75	0.0999025588907083\\
2.9	0.11544218912312\\
3.05	0.133371740603567\\
3.2	0.154156692729391\\
3.35	0.178356623812482\\
3.5	0.206642637287702\\
3.65	0.239818273884119\\
3.8	0.278844663866013\\
3.95	0.324870830244605\\
4.1	0.379270242662305\\
4.25	0.443684949408758\\
4.4	0.520078890227036\\
4.55	0.610802325556664\\
4.7	0.718669721251067\\
4.85	0.847053916961103\\
5	1\\
};
\addlegendentry{Magn. dipole ($\tau=1$), $\sigma=100$\unit{S/m}};

\addplot [color=red,solid,mark=o,mark options={solid}]
  table[row sep=crcr]{0.05	1.00443822309963\\
0.2	1.00443011296248\\
0.35	1.00441228605407\\
0.5	1.0043847770084\\
0.65	1.00434763934816\\
0.8	1.00430094548146\\
0.95	1.0042447866973\\
1.1	1.00417927316\\
1.25	1.0041045339024\\
1.4	1.00402071681796\\
1.55	1.00392798865174\\
1.7	1.00382653499021\\
1.85	1.00371656025007\\
2	1.00359828766586\\
2.15	1.00347195927658\\
2.3	1.00333783591127\\
2.45	1.00319619717351\\
2.6	1.00304734142495\\
2.75	1.00289158576787\\
2.9	1.00272926602673\\
3.05	1.00256073672888\\
3.2	1.00238637108429\\
3.35	1.00220656096444\\
3.5	1.0020217168804\\
3.65	1.00183226796002\\
3.8	1.00163866192447\\
3.95	1.00144136506396\\
4.1	1.00124086221278\\
4.25	1.00103765672369\\
4.4	1.00083227044173\\
4.55	1.0006252436774\\
4.7	1.00041713517933\\
4.85	1.00020852210651\\
5	1\\
};
\addlegendentry{El. dipole ($\tau=2$), $\sigma=1$\unit{S/m}};

\addplot [color=red,solid,mark=diamond,mark options={solid}]
  table[row sep=crcr]{0.05	0.916997168032768\\
0.2	0.916989985059085\\
0.35	0.916975570211942\\
0.5	0.916957045723023\\
0.65	0.916939236647248\\
0.8	0.916928670624023\\
0.95	0.916933577614335\\
1.1	0.916963889667863\\
1.25	0.917031240788318\\
1.4	0.917148966979347\\
1.55	0.917332106567448\\
1.7	0.917597400912443\\
1.85	0.917963295630133\\
2	0.918449942465874\\
2.15	0.919079201971973\\
2.3	0.919874647155851\\
2.45	0.920861568280154\\
2.6	0.922066979010172\\
2.75	0.923519624118151\\
2.9	0.925249988968447\\
3.05	0.927290311021843\\
3.2	0.929674593611858\\
3.35	0.932438622260561\\
3.5	0.935619983816129\\
3.65	0.93925808870945\\
3.8	0.94339419664221\\
3.95	0.948071446034413\\
4.1	0.953334887574991\\
4.25	0.959231522235315\\
4.4	0.965810344121859\\
4.55	0.973122388561205\\
4.7	0.981220785828049\\
4.85	0.990160820944797\\
5	1\\
};
\addlegendentry{El. dipole ($\tau=2$), $\sigma=10$\unit{S/m}};

\addplot [color=red,solid,mark=square,mark options={solid}]
  table[row sep=crcr]{0.05	0.0585326724258018\\
0.2	0.0585336254636153\\
0.35	0.0585445789084217\\
0.5	0.0585854623248105\\
0.65	0.0586870882470711\\
0.8	0.0588911897446234\\
0.95	0.059250509653976\\
1.1	0.0598289761137636\\
1.25	0.0607020082319823\\
1.4	0.0619570051234891\\
1.55	0.063694081337791\\
1.7	0.0660271220722992\\
1.85	0.0690852428302157\\
2	0.0730147507188857\\
2.15	0.0779817188797555\\
2.3	0.0841753021989695\\
2.45	0.0918119422083372\\
2.6	0.101140632848086\\
2.75	0.112449447607769\\
2.9	0.126073563788116\\
3.05	0.142405062787704\\
3.2	0.161904838270594\\
3.35	0.185117009033892\\
3.5	0.212686313023523\\
3.65	0.245379056425994\\
3.8	0.2841083109171\\
3.95	0.329964197577119\\
4.1	0.384250273235325\\
4.25	0.448527250790665\\
4.4	0.524665547487249\\
4.55	0.614908474042399\\
4.7	0.721948264877581\\
4.85	0.849017620002004\\
5	1\\
};
\addlegendentry{El. dipole ($\tau=2$), $\sigma=100$\unit{S/m}};

\end{axis}
\end{tikzpicture}%
\end{center}
\vspace{-7mm}
\caption{Illustration of the skin effect in terms of the power ratio $S_{\tau 1}(k,r)/S_{\tau 1}(k,r_2)$ where $r$ is ranging from $0$ to $r_2$
and where $r_2=5$\unit{cm} is the exterior design radius. Here, $l=1$ and the magnetic and electric dipole fields are indicated with $\tau=1$
and $\tau=2$, respectively.}
\label{fig:matfig52}
\end{figure}
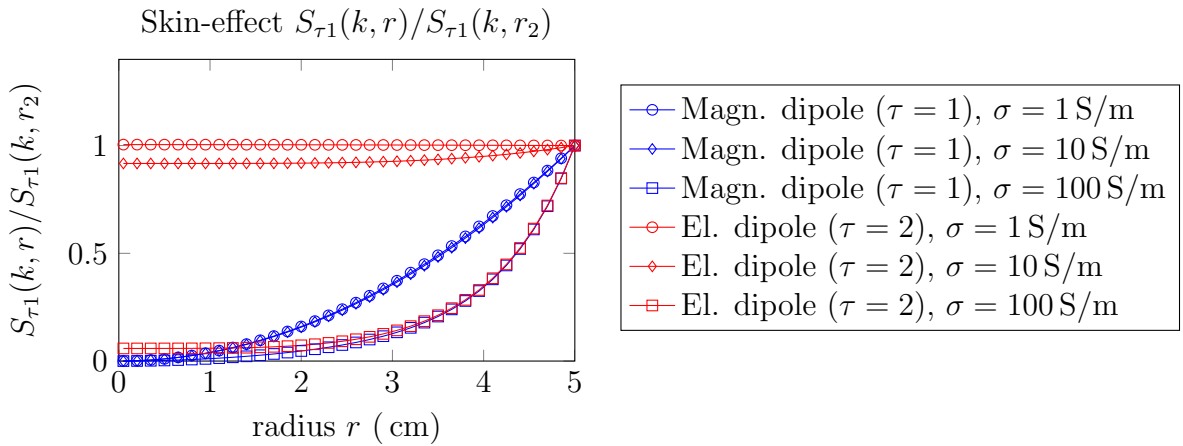

In Figure \ref{fig:matfig51} is shown the
relative heating $P_{\rm loc}(r_1)/P_{\rm b}(r_2)$ (in logarithmic scale) for gold nanoparticles with radius $r_1$ ranging from $0.8$\unit{nm} to
$25$\unit{nm} with a volume fraction of $f_1=0.01$ and with exterior design radius $r_2=5$\unit{cm}. The optimization in \eqref{eq:Plocr1Pbr2} yielded a maximum
for $l=1$ and the plot shows the resulting maximum with magnetic and electric dipole fields indicated with $\tau=1$
and $\tau=2$, respectively.  As seen in Figure \ref{fig:matfig51}, the relative heating $P_{\rm loc}(r_1)/P_{\rm b}(r_2)$ is extremely small and
can not give a practically useful heating, neighter with electric dipole effects nor with magnetic dipoles (inductive heating).

The reason that the latter conclusion is in contrast to the previous indications based on quasi-static assumptions and lossless Mie theory is that
the skin effect of the exterior bulk material must also be taken into account.
In Figure \ref{fig:matfig52} is illustrated the skin effect in terms of the power ratio $S_{\tau 1}(k,r)/S_{\tau 1}(k,r_2)$ defined
by \eqref{eq:Stauldef} and where $r$ is ranging from $0$ to $r_2$ and where $r_2=5$\unit{cm} is the exterior design radius in this example. 
It is seen that the skin effect is significant and destroys the possibility of radio frequency heating of gold nanoparticles in this example. 
In particular, with the magnetically induced dipoles and inductively heated nanoparticles ($\tau=1$)
the local heating of the particles may be significant but only to the price of an excessive heating in the skin of the exterior domain.

\section{Future research}
Experimental results performed at Middlesex University 
indicate that gold nanoparticles adhered with glutathiones can give
a significant heating effect when the antenna loaded with the specimen under test is tuned to resonance within the $2.6$\unit{GHz} band, \cf also \cite{marquez2013hyperthermia,Callaghan+etal2010}.
It is suggested that the glutathiones serving as electron donors may generate bound charges adhered 
to the gold nanoparticles and in this way providing a dielectric relaxation mechanism that is able to release significant losses,
\cf \eg the electrophoretic mechanisms that are due to the movement of nanoparticles having net charges \cite{Collins+etal2014,Sassaroli+etal2012}.
It is the aim of our future research to study the physical background of this effect and to derive useful dispersion and 
absorption models that can be used to analyze the related physical phenomena, see also \cite{Nordebo+etal2016a}. 

Gold is important as it can be hydrophilic and allow ligands to be attach to it. This will hence allow the GNPs 
to be targeted at the cancer cells. It has been found that a frequency of 2.6\unit{GHz} works well with the type of GNPs used, 
however different ligands will require different frequencies. Hence, there is a need to develop a model that allows us to predict the 
suitable frequency  instead of finding it experimentally. It is also important to find the best frequency that minimizes 
the heating of normal cells. 

\begin{figure}[htb]
\begin{picture}(50,140)
\put(150,0){\makebox(150,130){\includegraphics[width=12cm]{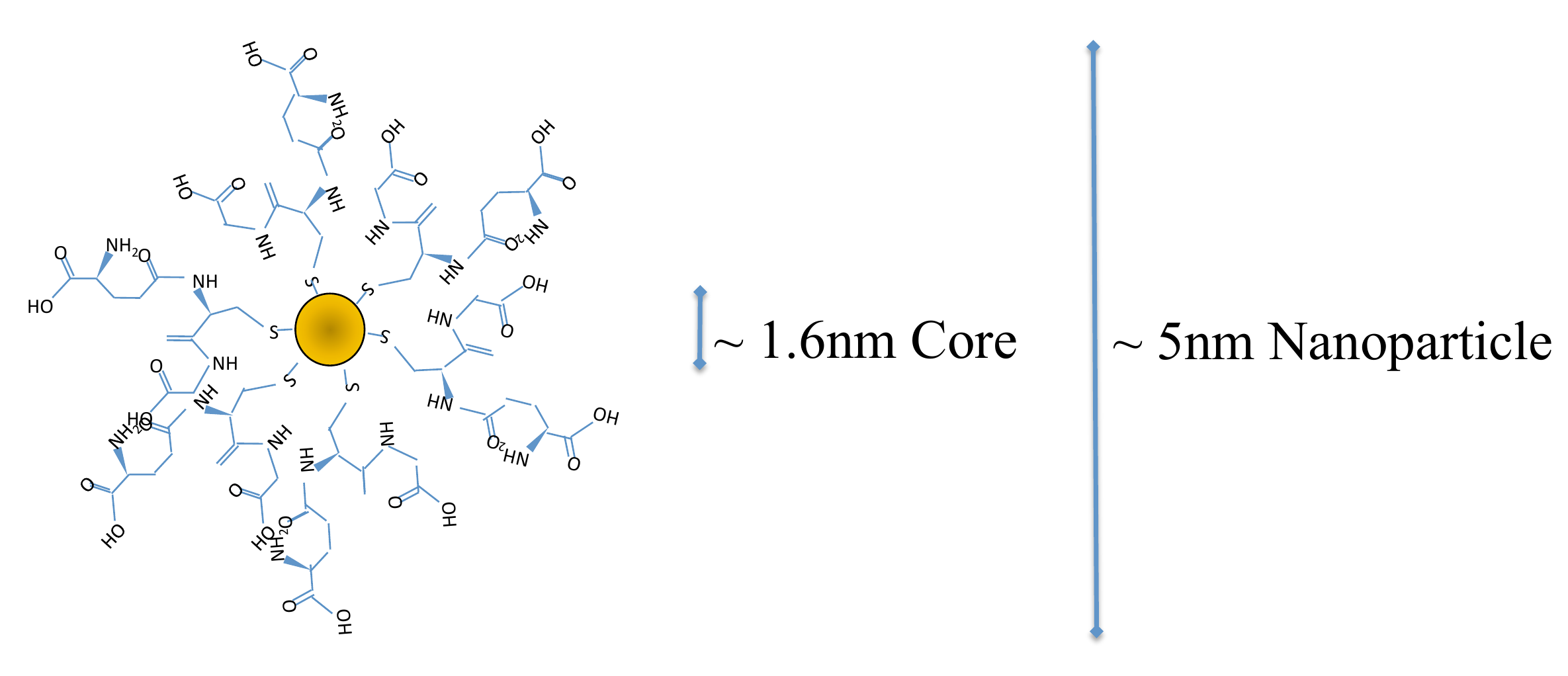}}} 
\end{picture}
\caption{Glutathione coated gold nanoparticle.}
\label{fig:HCPcell0}
\end{figure}

An illustration of the glutathione coated gold nanoparticle is shown in Figure \ref{fig:HCPcell0}.
Here, the structure consists of a 1.6\unit{nm} core of approximately 102 gold atoms. These GNPs
are coated with approximately 43 glutathione (GSH) ligands such that the total GSH-GNP diameter approaches 5\unit{nm}  \cite{Curley+etal2008,Callaghan+etal2010}.

Presently, the following issues are of great interest for future research
\begin{itemize}
\item Determination of dispersion models for the glutathione coated gold nanoparticles.
\begin{itemize}
\item[$\diamond$] This will involve a study of the various dispersion models based on nanoscale particles
that are available in the literature, such as \eg the electrophoretic mechanisms, nonlocal electronic surface effects, electron spill-out and surface roughness
that is proposed in \cite{Collins+etal2014,Hanson+etal2011,Sassaroli+etal2012}.  The aim is to be able to propose a plausible dispersion model and
an effective dielectric function for a nanoparticle suspension that is coherent with the observed experiments and that comprises
physical parameters that can be useful in the practical design of the nanoparticles as well as of the electronic heating system.
\item[$\diamond$] The evaluation of plausible dispersion models will be carried out based
on a careful analysis of the losses generated in the coated nanospheres together with an optimal near field generation including the 
associated skin effect and losses in the bulk exterior medium.
This will involve an extension and a generalization of the analysis that has been carried out in this paper 
based on the vector spherical waves, its associated energy expressions and optimization.
Our first results with examples including the electrophoretic mechanism are reported in \cite{Nordebo+etal2016a}.
\end{itemize}
\item Physical limitations for radio frequency absorption in gold nanoparticle suspensions.
\begin{itemize}
\item[$\diamond$] 
Given the physical dimensions regarding the conglomerated gold-glutathione nanoparticle as well as the proposed bandwidth of the RF-heating system,
it would be very useful being able to determine the physical limitations on the absorption that can be achieved by the particle, 
regardless of the passive dispersion model that is used.
Similar studies have recently been conducted in the context of passive metamaterials \cite{Gustafsson+Sjoberg2010a},
radar absorbers \cite{Rozanov2000}, high-impedance surfaces \cite{Gustafsson+Sjoberg2011},
as well as with antennas and scattering (extinction cross section, absorption efficiency, reflection coefficients, optical theorem and sum rules for periodic structures, etc.), 
see \eg \cite{Gustafsson+etal2010a,Gustafsson+etal2013a}.
In this context, physical bounds and associated sum rules have been derived successfully based on the assumptions of linearity, passivity, and causality using 
analytic function theory for Herglotz functions (or Positive Real (PR) functions) and some a priori knowledge regarding
the low-frequency asymptotics of the scattering bodies. 
The theoretical challenge here is that the particles are immersed in a lossy background, implying that many of the concepts used in normal scattering 
theory need to be adapted. Another option is to directly employ the passivity assumptions and related
representation formulas (such as the Hilbert transform) in the context of convex optimization \cite{Nordebo+etal2014b}.
Our first results on the physical limitations for radio frequency absorption in gold nanoparticle suspensions are reported in \cite{Nordebo+etal2016a}.
These results are constrained to spherical geometries and future research will therefore be devoted to investigate more general geometries of the nanoparticle suspension
and measurement set-ups.

\end{itemize}

\item Magnetic nanoparticles.
\begin{itemize}
\item[$\diamond$] The theoretical investigations conducted above will also involve the possibility to include magnetic properties.
Gold nanoparticles is the first choice due to their well-established bio-compatibility and their ability to be adhered with the cancer specific nutrients \cite{Dreaden+etal2012}.
However, it has been suggested that gold nanoparticles may exhibit magnetic properties \cite{Nealon+etal2012}, and the possibility to employ
ferromagnetic nanoparticles must also be kept open \cite{Hergt+etal2006}.
There are several mechanisms that have been proposed to account for magnetic losses such as Brownian or
viscous heating \cite{Collins+etal2014}.
In \cite{Collins+etal2014} is also proposed that magnetism in the sub-10\unit{nm} GNPs can give a significant heating effect
possibly in combination with the electrophoretic mechanisms which are due the movement of GNPs with net charges.
\end{itemize}

\item Antenna design based on predicted heating.
\begin{itemize}
\item[$\diamond$] This will involve electromagnetic design of both capacitively as well as magnetically coupled antennas depending on further experimental results
and theoretical findings. It will also involve thermodynamical analysis to determine the required antenna power and time for heating.
\end{itemize}
\end{itemize}

\section{Summary and conclusions}


We have investigated the classical electromagnetic background to the generation of heat in a cluster of conductive nanoparticles that are immersed in a lossy medium.
First, a quasi-electrostatic homogenization approach (Hashin-Shtrikman coated spheres) has been used to show that the effect of generating electric dipoles and the associated
losses dissipated in the exterior medium can be disregarded, unless the volume fraction of the GNPs is unrealistically high or if there are some other electric dipole mechanisms present
(such as \eg with nanospheres coated with ligands and associated electrophoretic mechanisms) which are not taken into account here. A quasi-magnetostatic analysis has then been performed
indicating that an inductive heating (induced eddy currents inside the metal particles) based on magnetic coupling
may have the potential to significantly increase the heating locally provided that the supplied magnetic field can be made sufficiently strong at radio frequency.  
Finally, a near field optimization approach has been presented to study the electromagentic heating of conductive nanoparticles 
showing that when the exterior medium is modelled as salty water the skin effect in the bulk material will render the simple principle of inductive heating of GNPs
practically useless at $13.56$\unit{MHz}. Future research will therefore be focused on an investigation of plausible dispersion mechanisms associated with GNPs coated with ligands and
which can potentially generate significant absorption based on the electrophoretic (plasmonic) resonance effects. The glutathione coated GNPs provide a promising candidate where experiments have indicated resonances in the GHz regime. To this end, it is anticipated that the presented near field optimization approach can be extended to provide a useful analysis tool
based on vector spherical waves for lossy materials and an optimization of the associated energy expressions.
Our first results in this direction with examples including the electrophoretic mechanism are reported in \cite{Nordebo+etal2016a}.


\appendix
\section{Vector spherical waves}\label{sect:spherical}
\subsection{Definition of vector spherical waves}
In a source-free region the electromagnetic field can be expanded in vector spherical waves (multipoles) as
\begin{equation}\label{eq:Esphdef}
\left\{\begin{array}{l}
\vec{E}(\vec{r})=\displaystyle\sum_{l=1}^{\infty}\sum_{m=-l}^{l}\sum_{\tau=1}^2a_{\tau ml}\vec{v}_{\tau ml}(k\vec{r})+b_{\tau ml}\vec{u}_{\tau ml}(k\vec{r}), \vspace{0.2cm} \\
\vec{H}(\vec{r})=\displaystyle\frac{1}{\iu\eta_0\eta}\sum_{l=1}^{\infty}\sum_{m=-l}^{l}\sum_{\tau=1}^2a_{\tau ml}\vec{v}_{\bar{\tau} ml}(k\vec{r})+b_{\tau ml}\vec{u}_{\bar{\tau} ml}(k\vec{r}),
\end{array}\right.
\end{equation}
where $\vec{v}_{\tau ml}(k\vec{r})$ and $\vec{u}_{\tau ml}(k\vec{r})$ are the regular and the outgoing vector spherical waves, respectively, 
\cf \cite{Bostrom+Kristensson+Strom1991,Arfken+Weber2001,Jackson1999,Newton2002}.
Here, $\tau=1$ indicates a transverse electric (\textrm{TE}) magnetic multipole and $\tau=2$ a transverse magnetic (\textrm{TM}) electric multipole,
and $\bar{\tau}$ denotes the complement of $\tau$.

The solenoidal (source-free) regular vector spherical waves are defined here by
\begin{equation}\label{eq:vdef}
\left\{\begin{array}{l}
\displaystyle\vec{v}_{1 ml}(k{\bm{r}})  =   \frac{1}{\sqrt{l(l+1)}}\nabla\times({\vec{r}}{\rm j}_l(kr)Y_{ml}(\hat{\vec{r}})) =   {\rm j}_l(kr)\vec{A}_{1 ml}(\hat{\vec{r}}), \vspace{0.2cm}\\
\vec{v}_{2 ml}(k\vec{r})   =   \displaystyle \frac{1}{k}\nabla\times\vec{v}_{1 ml}(k\vec{r})
 =\displaystyle\frac{(kr{\rm j}_l(kr))^{\prime}}{kr}\vec{A}_{2 ml}(\hat{\vec{r}})+\sqrt{l(l+1)}\frac{{\rm j}_l(kr)}{kr}\vec{A}_{3 ml}(\hat{\vec{r}}), 
\end{array}\right.
\end{equation}
where $Y_{ml}(\hat{\vec{r}})$ are the spherical harmonics, $\bm{A}_{\tau ml}(\hat{\bm{r}})$ the vector spherical harmonics and ${\rm j}_l(x)$ the spherical Bessel functions of order $l$,
\cf \cite{Bostrom+Kristensson+Strom1991,Arfken+Weber2001,Jackson1999,Newton2002,Olver+etal2010}. 
Note that the indices $l$ and $m$ are given by $l=1,\ldots,\infty$ and $m=-l,\ldots,l$.
The outgoing (radiating) vector spherical waves $\vec{u}_{\tau ml}(k{\vec{r}})$ are obtained by replacing
the regular spherical Bessel functions ${\rm j}_l(x)$ above for the spherical Hankel functions of the first kind, ${\rm h}_l^{(1)}(x)$, see \cite{Bostrom+Kristensson+Strom1991,Olver+etal2010}.
It can be shown that any one of the vector spherical waves $\vec{w}_{\tau ml}(k\vec{r})$ defined above satisfy the following curl properties
\begin{equation}\label{eq:w12cross}
\nabla\times \vec{w}_{\tau ml}(k\vec{r})=k\vec{w}_{\bar{\tau} ml}(k\vec{r}),
\end{equation}
and hence the source-free Maxwell's equations (vector Helmholtz equation) in free space, \ie 
$\nabla\times\nabla\times \vec{w}_{\tau ml}(k\vec{r})=k^2 \vec{w}_{\tau ml}(k\vec{r})$.

The vector spherical harmonics $\bm{A}_{\tau lm}(\hat{\bm{r}})$ are given by
\begin{equation}\label{eq:Adef}
\left\{\begin{array}{l}
\vec{A}_{1ml}(\hat{\vec{r}})  =   \displaystyle\frac{1}{\sqrt{l(l+1)}}\nabla\times\left( \vec{r}{\rm Y}_{ml}(\hat{\vec{r}}) \right), \vspace{0.2cm}\\
\vec{A}_{2ml}(\hat{\vec{r}})  =  \hat{\vec{r}}\times\bm{A}_{1ml}(\hat{\vec{r}}), \vspace{0.2cm}\\
\vec{A}_{3ml}(\hat{\vec{r}}) = \hat{\vec{r}}{\rm Y}_{ml}(\hat{\vec{r}}),
\end{array}\right.
\end{equation}
where the spherical harmonics ${\rm Y}_{ml}(\hat{\vec{r}})$ are given by
\begin{equation}
Y_{ml}(\hat{\vec{r}})=(-1)^m\sqrt{\frac{2l+1}{4\pi}}\sqrt{\frac{(l-m)!}{(l+m)!}}{\rm P}_{l}^m(\cos\theta)\eu^{{\rm i}m\phi},
\end{equation}
and where ${\rm P}_{l}^m(x)$ are the associated Legendre functions \cite{Arfken+Weber2001,Olver+etal2010}.
Important symmetry properties are $P_l^{-m}(\cos\theta)=(-1)^m\frac{(l-m)!}{(l+m)!}P_l^m(\cos\theta)$ and 
$Y_{-m,l}(\theta,\phi)=(-1)^mY_{ml}^*(\theta,\phi)$.
The vector spherical harmonics are orthonormal on the unit sphere, and hence
\begin{equation}\label{eq:Aorthonormal}
\int_{\Omega}\vec{A}_{\tau ml}^*(\hat{\vec{r}})\cdot\vec{A}_{\tau^\prime m^\prime l^\prime }(\hat{\vec{r}})\diff\Omega
=\delta_{\tau\tau^\prime}\delta_{mm^\prime}\delta_{ll^\prime},
\end{equation}
where $\Omega$ denotes the unit sphere, $\diff\Omega=\sin\theta\diff\theta\diff\phi$ and $\tau=1,2,3$. 

In spherical coordinates the vector spherical harmonics are given by
\begin{equation}
\left\{\begin{array}{l}
\vec{A}_{1ml}(\hat{\vec{r}})=\displaystyle\frac{1}{\sqrt{l(l+1)}}\left(\hat{\vec{\theta}}\frac{1}{\sin\theta}\frac{\partial}{\partial \phi}Y_{ml}(\hat{\vec{r}})
-\hat{\vec{\phi}}\frac{\partial}{\partial \theta}Y_{ml}(\hat{\vec{r}}) \right), \vspace{0.2cm} \\
\vec{A}_{2ml}(\hat{\vec{r}})=\displaystyle\frac{1}{\sqrt{l(l+1)}}\left(\hat{\vec{\theta}}\frac{\partial}{\partial \theta}Y_{ml}(\hat{\vec{r}})
+\hat{\vec{\phi}}\frac{1}{\sin\theta}\frac{\partial}{\partial \phi}Y_{ml}(\hat{\vec{r}})\right),\vspace{0.2cm} \\
\vec{A}_{3ml}(\hat{\vec{r}})=\displaystyle\hat{\vec{r}}Y_{ml}(\hat{\vec{r}}).
\end{array}\right.
\end{equation}

An important special case is with $(m,l)=(0,1)$ corresponding to a dipole moment in the $\hat{\vec{z}}$-direction, where $Y_{01}(\hat{\vec{r}})=\sqrt{\frac{3}{4\pi}}\cos\theta$
and
\begin{equation}
\left\{\begin{array}{l}
\vec{A}_{101}(\hat{\vec{r}})=\sqrt{\frac{3}{8\pi}}\hat{\vec{\phi}}\sin\theta, \vspace{0.2cm} \\
\vec{A}_{201}(\hat{\vec{r}})=-\sqrt{\frac{3}{8\pi}}\hat{\vec{\theta}}\sin\theta, \vspace{0.2cm} \\
\vec{A}_{301}(\hat{\vec{r}})=\sqrt{\frac{3}{4\pi}}\hat{\vec{r}}\cos\theta.
\end{array}\right.
\end{equation}

\subsection{Lommel integrals for spherical Bessel functions}
The two Lommel integrals  are
\begin{equation}\label{eq:Lommel1}
\int {\rm C}_{\nu}(a\rho){\rm D}_\nu(b\rho)\rho\diff\rho=
\frac{\rho\left(a{\rm C}_{\nu+1}(a\rho) {\rm D}_{\nu}(b\rho) -b{\rm C}_{\nu}(a\rho) {\rm D}_{\nu+1}(b\rho) \right)}{a^2-b^2},
\end{equation}
and
\begin{equation}\label{eq:Lommel2}
\int {\rm C}_{\nu}(a\rho){\rm D}_{\nu}(a\rho)\rho\diff\rho=\frac{1}{4}\rho^2\left(2{\rm C}_{\nu}(a\rho){\rm D}_{\nu}(a\rho)
-{\rm C}_{\nu-1}(a\rho){\rm D}_{\nu+1}(a\rho)-{\rm C}_{\nu+1}(a\rho){\rm D}_{\nu-1}(a\rho) \right),
\end{equation}
where $a$ and $b$ are complex valued constants and ${\rm C}_{\nu}(\cdot)$ and ${\rm D}_{\nu}(\cdot)$ are arbitrary cylinder functions, \ie 
the Bessel function, the Neumann function, the Hankel functions of the first and second kind
${\rm J}_{\nu}(\cdot)$, ${\rm Y}_{\nu}(\cdot)$, ${\rm H}_{\nu}^{(1)}(\cdot)$, ${\rm H}_{\nu}^{(2)}(\cdot)$, respectively, or any nontrivial linear combination
of these functions, see 10.22.4 and 10.22.5 in \cite{Olver+etal2010},  and pp.\ 133--134 in \cite{Watson1995}.

Let $a=\kappa$ and $b=\kappa^*$ where $\kappa\neq\kappa^*$, \ie $\kappa$ is not real valued, and consider the case
\begin{equation}\label{eq:CLomdef}
{\rm C}_\nu(\kappa\rho)=A {\rm J}_\nu(\kappa\rho)+B{\rm H}_\nu^{(1)}(\kappa\rho),
\end{equation}
where $A$ and $B$ are complex valued constants. Let
\begin{equation}\label{eq:DLomdef1}
{\rm D}_{\nu}(\kappa^*\rho)={\rm C}_\nu^*(\kappa\rho)=A^* {\rm J}_\nu(\kappa^*\rho)+B^*{\rm H}_\nu^{(2)}(\kappa^*\rho),
\end{equation}
where the conjugate rules ${\rm J}_\nu^*(\zeta)={\rm J}_\nu(\zeta^*)$ and ${{\rm H}_\nu^{(1)}}^*(\zeta)={\rm H}_\nu^{(2)}(\zeta^*)$
have been used, see \cite{Olver+etal2010}.
The first Lommel integral \eqref{eq:Lommel1} now yields
\begin{equation}\label{eq:Lommel1b}
\int |{\rm C}_\nu(\kappa\rho)|^2\rho\diff\rho=
\frac{\rho\Im\left\{\kappa{\rm C}_{\nu+1}(\kappa\rho) {\rm C}_\nu^*(\kappa\rho) 
 \right\}}{\Im\left\{\kappa^2\right\}}.
\end{equation}

The spherical Bessel, Neumann and Hankel functions of the first and second kind are given by
${\rm j}_l(\zeta)=\sqrt{\frac{\pi}{2\zeta}}{\rm J}_{l+1/2}(\zeta)$, ${\rm y}_l(\zeta)=\sqrt{\frac{\pi}{2\zeta}}{\rm Y}_{l+1/2}(\zeta)$,
${\rm h}_l^{(1)}(\zeta)=\sqrt{\frac{\pi}{2\zeta}}{\rm H}_{l+1/2}^{(1)}(\zeta)$ and ${\rm h}_l^{(2)}(\zeta)=\sqrt{\frac{\pi}{2\zeta}}{\rm H}_{l+1/2}^{(2)}(\zeta)$, 
respectively, see \cite{Olver+etal2010}. An arbitrary linear combination of spherical Bessel and Hankel functions can hence be written as
\begin{equation}
{\rm s}_l(kr)=A {\rm j}_l(kr)+B{\rm h}_l^{(1)}(kr)=\sqrt{\frac{\pi}{2kr}}{\rm C}_{l+1/2}(kr),
\end{equation}
where ${\rm C}_{l+1/2}(kr)$ is the corresponding cylinder function as defined in \eqref{eq:CLomdef}. 
The first Lommel integral for spherical Bessel functions  with complex valued arguments can now be derived as
\begin{multline}\label{eq:firstLommelcomplxarg}
\int\left|{\rm s}_l(kr)\right|^2r^2\diff r=
\int\frac{\pi}{2|kr|}\left|{\rm C}_{l+1/2}(kr) \right|^2r^2\diff r
=\frac{\pi}{2|k|}\int\left|{\rm C}_{l+1/2}(kr) \right|^2r\diff r \\
=\frac{\pi}{2|k|}\frac{r\Im\left\{k{\rm C}_{l+1+1/2}(kr) {\rm C}_{l+1/2}^*(kr) \right\}}{\Im\left\{k^2\right\}}
=\frac{r^2 \Im\left\{k\sqrt{\frac{\pi}{2kr}}{\rm C}_{l+1+1/2}(kr) \left(\sqrt{\frac{\pi}{2kr}}{\rm C}_{l+1/2}(kr)\right)^* \right\}}{\Im\left\{k^2\right\}} \\
=\frac{r^2 \Im\left\{k{\rm s}_{l+1}(kr) {\rm s}_{l}^*(kr) \right\}}{\Im\left\{k^2\right\}}.
\end{multline}

Consider next the function ${\rm C}_\nu(\kappa\rho)$ defined as in \eqref{eq:CLomdef} for the case when $a=\kappa$ is real valued.
In this case we have
\begin{equation}\label{eq:DLomdef2}
{\rm D}_{\nu}(\kappa\rho)={\rm C}_\nu^*(\kappa\rho)=A^* {\rm J}_\nu(\kappa\rho)+B^*{\rm H}_\nu^{(2)}(\kappa\rho),
\end{equation}
and the second Lommel integral \eqref{eq:Lommel2} yields
\begin{equation}\label{eq:Lommel2b}
\int |{\rm C}_\nu(\kappa\rho)|^2\rho\diff\rho=
\frac{1}{2}\rho^2\left( \left| {\rm C}_{\nu}(\kappa\rho)\right|^2-\Re\{{\rm C}_{\nu-1}(\kappa\rho){\rm C}_{\nu+1}^*(\kappa\rho)\} \right).
\end{equation}
The second Lommel integral for spherical Bessel functions with real valued arguments can now be derived as
\begin{multline}\label{eq:secondLommelcomplxarg}
\int\left|{\rm s}_l(kr)\right|^2r^2\diff r=
\int\frac{\pi}{2kr}\left|{\rm C}_{l+1/2}(kr) \right|^2r^2\diff r
=\frac{\pi}{2k}\int\left|{\rm C}_{l+1/2}(kr) \right|^2r\diff r \\
=\frac{\pi}{2k}\frac{1}{2}r^2\left( \left| {\rm C}_{l+1/2}(kr)\right|^2-\Re\{{\rm C}_{l-1+1/2}(kr){\rm C}_{l+1+1/2}^*(kr)\} \right) \\
=\frac{1}{2}r^3\left( \left| \sqrt{\frac{\pi}{2kr}}{\rm C}_{l+1/2}(kr)\right|^2-\Re\{\sqrt{\frac{\pi}{2kr}}{\rm C}_{l-1+1/2}(kr)
\sqrt{\frac{\pi}{2kr}}{\rm C}_{l+1+1/2}^*(kr)\} \right) \\
=\frac{1}{2}r^3\left( \left| {\rm s}_{l}(kr)\right|^2-\Re\{{\rm s}_{l-1}(kr){\rm s}_{l+1}^*(kr)\} \right).
\end{multline}

\subsection{Orthogonality of the regular spherical waves}
Due to the orthonormality of the vector spherical harmonics \eqref{eq:Aorthonormal}
the regular spherical waves are orthogonal over the unit sphere with
\begin{equation}\label{eq:vorthogonal1}
\displaystyle\int_{\Omega}\vec{v}_{\tau ml}^*(k{\vec{r}})\cdot\vec{v}_{\tau^\prime m^\prime l^\prime}(k{\vec{r}})\diff \Omega
=\displaystyle\delta_{\tau\tau^\prime}\delta_{mm^\prime}\delta_{ll^\prime}S_{\tau l}(k,r),
\end{equation}
where
\begin{equation}\label{eq:Stauldef}
S_{\tau l}(k,r)=\displaystyle\int_{\Omega}|\vec{v}_{\tau ml}(k\bm{r})|^2\diff\Omega
=\left\{\begin{array}{ll}
\displaystyle  \left|{\rm j}_{l}(kr)\right|^2 & \textrm{for}\  \tau=1,  \vspace{0.2cm} \\
\displaystyle  \left|\frac{{\rm j}_{l}(kr)}{kr}+{\rm j}_{l}^\prime(kr)\right|^2+l(l+1) \left|\frac{{\rm j}_{l}(kr)}{kr}\right|^2  & \textrm{for}\  \tau=2.
\end{array}\right.
\end{equation}
As a consequence, the regular spherical waves are also orthogonal over a spherical volume $V_{r_1}$ with radius $r_1$ yielding
\begin{equation}\label{eq:vorthogonal2}
\displaystyle\int_{V_{r_1}}\vec{v}_{\tau ml}^*(k{\vec{r}})\cdot\vec{v}_{\tau^\prime m^\prime l^\prime}(k{\vec{r}})\diff v
=\displaystyle\delta_{\tau\tau^\prime}\delta_{mm^\prime}\delta_{ll^\prime}W_{\tau l}(k,r_1),
\end{equation}
where
\begin{equation}\label{eq:Wtauldef}
W_{\tau l}(k,r_1)=\int_{V_{r_1}}\left|\vec{v}_{\tau ml}(k\bm{r})\right|^2\diff v=\int_{0}^{r_1}S_{\tau l}(k,r)r^2\diff r,
\end{equation}
where $\diff v=r^2\diff\Omega\diff r$ and $\tau=1,2$.

For complex valued arguments $k$, $W_{1l}(k,r_1)$ is obtained from \eqref{eq:firstLommelcomplxarg} as
\begin{equation}\label{eq:W1ldef1}
W_{1l}(k,r_1)=\int_{0}^{r_1}\left|{\rm j}_{l}(kr)\right|^2 r^2\diff r
=\frac{r_1^2 \Im\left\{k{\rm j}_{l+1}(kr_1) {\rm j}_{l}^*(kr_1) \right\}}{\Im\left\{k^2\right\}},
\end{equation}
and for real valued arguments $k$, $W_{1l}(k,r_1)$  is obtained from \eqref{eq:secondLommelcomplxarg} as
\begin{equation}\label{eq:W1ldef2}
W_{1l}(k,r_1)=\int_{0}^{r_1}{\rm j}_{l}^2(kr) r^2\diff r
=\frac{1}{2}r_1^3\left(  {\rm j}_{l}^2(kr_1)-{\rm j}_{l-1}(kr_1){\rm j}_{l+1}(kr_1) \right).
\end{equation}
By using the following recursive relationships
\begin{equation}
\left\{\begin{array}{l}
\displaystyle \frac{{\rm j}_l(kr)}{kr}=\frac{1}{2l+1}\left({\rm j}_{l-1}(kr)+{\rm j}_{l+1}(kr) \right) \vspace{0.2cm} \\
\displaystyle {\rm j}_l^\prime(kr)=\frac{1}{2l+1}\left(l{\rm j}_{l-1}(kr)-(l+1){\rm j}_{l+1}(kr) \right),
\end{array}\right.
\end{equation}
which are valid for $l=1,2,\ldots$, \cf \cite{Olver+etal2010}, it is straightforward to show that
\begin{multline}\label{eq:W2ldef}
W_{2 l}(k,r_1)=\int_{0}^{r_1}\left(\left|\frac{{\rm j}_{l}(kr)}{kr}+{\rm j}_{l}^\prime(kr)\right|^2+l(l+1) \left|\frac{{\rm j}_{l}(kr)}{kr}\right|^2\right)r^2\diff r \\
=\frac{1}{2l+1}\left((l+1)W_{1,l-1}(k,r_1)+lW_{1,l+1}(k,r_1) \right).
\end{multline}



\end{document}